\documentstyle[11pt,psfig]{article}       
\begin{document}

\begin{flushright}
THU-97/39
\end{flushright}
\begin{center}
{\large\bf
Analytic profiles and correlation functions in\\
synchronous asymmetric exclusion processes}

\vspace{1cm}

L.G. Tilstra \footnote{e-mail: tilstra@fys.ruu.nl} and M.H. Ernst \footnote{e-mail: ernst@fys.ruu.nl}

\vspace{1cm}

{\it Institute for Theoretical Physics\\
University of Utrecht\\
Princetonplein 5, 3584 CC Utrecht\\
The Netherlands}
\end{center}
\vspace{1cm}

\date{\today}

\begin{abstract}
\noindent
The transfer matrix and matrix multiplication ansatz, when applied to 
nonequilibrium steady states in asymmetric exclusion processes and 
traffic models, has given many exact results for phase diagrams, bulk 
densities and fluxes, as well as density profiles and spatial and 
temporal correlation functions for models with a dynamics that is 
updated in (random) sequential and sublattice-parallel order. 
Here we consider fully parallel or synchronous dynamics, for which 
only partial results are known, due to the appearance of complicating 
strong short range correlations, that invalidate simple mean field 
approximations.\\
This paper is based on two new ingredients: (i) a microscopic 
characterization of order parameters and local configurations in the 
relevant phases, based on the microdynamics of the model, and (ii) 
 an improved mean field approximation, which neglects certain four 
point -- and higher order correlation functions. It is
{\it conjectured} that the density profiles, obtained here, are 
exact up to terms that are exponentially small in the system size.
\end{abstract}


\newpage

\section{Introduction}

Nonequilibrium stationary states (NESS)  violate 
detailed balance, they cannot be described as Gibbs states, and
their behavior shows a wealth of interesting phenomena that are 
absent in thermal equilibrium, such as  boundary induced phase 
transitions, self organization, pattern formation, and long range 
spatial and temporal correlations. They occur in classical fluids 
\cite{DKS}, driven diffusive systems \cite{schmittmann,spohn,grinstein}, 
granular flows \cite{vnoije-granflow} and
lattice gas cellular automata (LGCA) with collision rules violating 
detailed balance \cite{harmen}, of which traffic models \cite{rsss8} 
are simple examples. Unfortunately, a general theory like Gibbs 
statistical mechanics is lacking for NESS, where results seem much less 
universal, and depend strongly on boundary conditions, driving forces, and 
the (sequential or synchronous) order in which the  microscopic dynamics 
is applied \cite{rsss}.

The standard theoretical approaches are based on Langevin equations, 
fluctuating hydrodynamics, mode coupling theories, and ring kinetic 
theory, which are phenomenological and/or approximate in nature. 
A large amount of theoretical understanding has also been 
obtained from computer simulations as well.

However since 1992 new methods for obtaining exact solutions for 
simple open one-dimensional systems, 
the so-called asymmetric exclusion processes, have been developed 
based on transfer matrix methods (Bethe ansatz, matrix product ansatz
 \cite{derrida}-\cite{schreck}). One can calculate bulk properties, 
phase diagrams, density profiles of boundary layers and interfaces 
between coexisting bulk phases, as well as spatial and temporal 
correlation functions. The number of exact results is rapidly 
growing. For an up to date list of references to analytic results 
for asymmetric exclusion processes we refer to the reader to Ref. \cite{rsss}. 

Asymmetric exclusion processes can be characterized as open systems of 
interacting particles or random walkers 
on a lattice, in most cases linear chains, with hard core exclusion 
for double occupancy, and with hopping rates differing for different 
directions.  
The bulk dynamics may be deterministic or stochastic. Open systems 
are coupled to reservoirs at both ends through stochastic boundaries.
Closed systems on a ring are coupled to local 
randomness such as defect sites \cite{schuetz-ring,yukawa} or 
defect particles \cite{mallick,evans,defect}.
Applications range from shock waves in the Burgers 
equation, to traffic flow problems, reaction-diffusion systems
and growth models.

The problem can be formulated in terms of master equations with 
discrete or continuous time \cite{derrida},
or in terms of the equivalent transfer matrices for spin chains 
\cite{stinchcombe}, 
or as microdynamic equations, as is usually done in LGCA, when used as 
models for nonequilibrium fluids \cite{6-gang}. LGCA represent perhaps 
also a more faithful representation of real traffic problems 
\cite{rsss8,evans}. The LGCA approach will be followed in this paper.

The dynamics of updating sites may be applied in (random) sequential
 order, typical for the master equation description, or in parallel, i.e. 
fully synchronous for all sites, typical for LGCA,
or in any intermediate version with strictly sequential or with 
sublattice--parallel updating \cite{schuetz,hinrichsen}. The different 
ways of updating are an essential part of the model. They affect 
the existence of different phases in the phase diagram, as well as the
structure of the spatial and temporal correlations.
For instance, the so-called maximum current phase is present in the 
totally asymmetric exclusion process 
(TASEP) with open boundaries, when updating is carried out in random 
sequential order \cite{derrida,domany}, but is absent for other updating 
schemes \cite{schuetz,hinrichsen,yukawa,tilstra}. The spatial 
correlations are weakest for random sequential updating, intermediate 
for sequential and sublattice--parallel updatings, and strongest 
for parallel updating.

As far as analytic approaches are concerned many exact results about 
bulk properties, spatial and temporal correlations, and profiles 
are known for random sequential and sublattice--parallel updatings 
\cite{rsss8,rsss,derrida,schuetz}. The fully parallel 
updating schemes of LGCA offer the largest difficulties, because the
dynamics creates strong short range correlations, which invalidate 
simple mean field approaches. Only some bulk properties like density, 
flux, and the phase diagram have been 
obtained \cite{rsss,yukawa,tilstra}, where `in the bulk' means 
`~outside the boundary layers'. 

The main goal of the present paper is an analytic 
derivation of the correlation functions  and the profiles,
conjectured in Ref. \cite{tilstra} for an open 
TASEP with parallel updating, where the particles move forward 
fully synchronously at 
every time step with probability $p=1$ (deterministic bulk dynamics) 
if the site in front of them is empty, and where input and removal rates 
specify the stochastic boundary conditions. In Ref. \cite{rsss}
a much richer stochastic version of the same model with $ p<1$ has 
been considered, and mean field results, based on a
matrix multiplication ansatz, have been obtained. It includes the 
deterministic TASEP, discussed in this paper, as a special case, and 
it also contains the maximal current phase, even in the case of fully 
synchronous updating. However, even with the 
highly sophisticated matrix multiplication ansatz, the profiles and 
correlation functions have not yet been calculated as the 
associated matrix algebra is quite complicated \cite{rsss}.
In case of sequential or sublattice updating the model of Ref. \cite{rsss}
has already been generalized to include backward jumps with probability 
$q$, and input and removal rates at both ends of the chain \cite{non-tasep}.

Our approach starts in section 2 from the {\it microdynamic equations}, 
which describe the time evolution of the set of occupation numbers 
$\{ \tau_i(t)\}$ of the sites $i=1,2, \cdots, L$ on the chain at time $t$, 
as a discrete continuity equation, in which all variables are Boolean 
variables, having only values 0 or 1:
\begin{equation} \label{a1}
\tau_i (t+1)  - \tau_i (t) = \hat{\jmath}_{i-1} (t)- \hat{\jmath}_{i} (t).
\end{equation} 
Here the instantaneous microscopic flux $\hat{\jmath}_{i} (t)$  counts the 
number of particles passing through link $(i,i+1)$ at time $t=0,1,2, \cdots$. 
The  influx $\hat{\jmath}_{0} (t)$ and outflux $\hat{\jmath}_{L} (t)$
specify the 
couplings to the stochastic reservoirs or blockage sites.  Further 
specification of the fluxes depends on the model, and will be given later.

A detailed analysis of the microscopic evolution  equation for sets 
of particle clusters $\tau_i \tau_{i+1} \cdots$, derived in section 3, 
allows us not only to calculate the short range correlation functions that are 
built up through the dynamics, but also to identify the different phases, 
as well as the microscopic structure of boundary layers and the interface 
between different bulk phases.
In section 4 the phase diagram, bulk densities and fluxes in the different 
phases are calculated.
The phase diagram  shows a phase transition from a free flow regime 
$(\alpha < \beta)$, where all particles are moving at maximum speed, to a 
congested or jammed regime $(\alpha > \beta)$, where the dynamics is 
controlled by start-stop waves. Here $\alpha$ is the input rate and 
$\beta$ the removal rate. When both rates are equal, there are coexisting 
phases with a sharp interface (shock wave) between them. The interpretation 
of the interface as a shock wave forms the direct link with Burgers
nonlinear diffusion equation \cite{joel-asep}.

In section 5 an exact hierarchy for 
particle-cluster correlation functions is derived, to which we apply our
 improved mean field approximation (MFA). It assumes for the low 
density phase, that {\it inside the interface} between bulk phase and 
boundary layer, the higher order correlations between on the one hand 
a particle-hole pair, and on the other hand the particle cluster 
directly in front of it, can be neglected. This is equivalent to neglecting
certain 4-point correlation functions.
For the high density phase, 
similar results are obtained through particle-hole symmetry. 
The MFA reduces the hierarchy to a set of recursion relations for 
the correlation functions and density profiles, which are solved 
in this section. Knowledge of the density profile also enables us to 
calculate the average travel time of particles.
The results are new, and are in excellent agreement with extensive 
computer simulations. The present MFA is similar in spirit, but not 
in specific details, to the improved MFA of Ref. \cite{rsss8}
that accounts for short range 2-point correlations, but neglects higher 
order ones, and that leads to the solution of the TASEP on a ring without 
a blockage.

To test the validity of our MFA we have applied the method in  the
appendix to the same open TASEP, but now with sublattice--parallel 
updating, for which the exact correlations functions have been 
calculated in Refs. \cite{schuetz,hinrichsen}. It appears that the results 
agree with the exact results, apart from terms that are 
exponentially small in the length of the chain. 
Moreover, it turns out that spatial correlations arising from parallel 
updating are quite different from those coming from
sublattice--parallel updating. 

In section 6 we exploit the analogy that a blockage site on a ring 
has with on the one hand the entrance site and on the other hand with 
the exit site of the corresponding open system. In this way we 
recover the results of Ref. \cite{yukawa} for the phase diagram, fluxes 
and the bulk densities for the TASEP with a blockage on a ring with
parallel updating. In addition we are able to
construct the higher order correlation functions, density profiles 
and finite size corrections (rounding) of the $j(\rho)$--relation 
(`equation of state') at the high and low density ends of the
coexisting phase region.
 Again, the results are new and in good agreement with computer 
simulations. We end in section 7 with some conclusions and suggestions.

\section{TASEP with synchronous dynamics}

\subsection{Definitions}

We consider a totally asymmetric exclusion process with open 
boundaries as a lattice gas cellular automaton, with particles 
living 
on a one-dimensional lattice with sites labeled $i = \{1,\cdots,L\}$. 
The configuration of particles at time $t=0,1,2,\cdots$ is described by the
set of occupation numbers $\{\tau_i(t)\}$ with $i = 1,2,\cdots,L$, where
$\tau_i(t) \equiv 1 - \sigma_i(t) = 1$ if the site $i$ is occupied by a
particle, and $\tau_i(t) = 0$ if the site is empty, i.e. is occupied by a
hole ($\sigma_i(t) = 1$).
The dynamics is defined such that all
particles, which have an {\em empty} nearest neighbor site in front 
of them at time $t$, simultaneously jump to that site at the next time 
step ($t+1$). If that site is occupied, the particle does not move.
So, the dynamics or updating in the bulk of the system is deterministic 
and fully parallel.

Next, {\em boundary conditions} are specified. We consider an open system,
coupled to two stochastic reservoirs, one that injects particles with a
probability $\alpha$ ($0<\alpha\leq 1$) into site 1, provided it is empty,
and one that removes particles from site $L$ with probability $\beta$
($0<\beta\leq 1$) provided site $L$ is occupied.

If the system is in configuration $\{\tau_i=\tau_i(t)\}$ at time $t$, then, 
according to the dynamic rules described above,
the configuration $\{\tau_i'=\tau_i(t+1)\}$ is  given by the microdynamic 
equation $\tau_i^\prime = \tau_i  + \hat{\jmath}_{i-1}- \hat{\jmath}_{i}$
with bulk and boundary fluxes given by\footnote{ In
the model of Ref. \cite{rsss} the boundary fluxes are the same, but the bulk 
flux is generalized to $\hat{\jmath}_{i} =$  $ \hat{p}_i \tau_{i} 
\sigma_{i+1}$, where $\hat{p}_{i}$ with $<\hat{p}_{i}> = p$ represents a set 
of independent Boolean variables, similar to 
$\hat{\alpha}$ and $\hat{\beta}$.} 
\begin{eqnarray} \label{a3}
\hat{\jmath}_{i} & = & \tau_{i}\sigma_{i+1} \qquad (i = 1, \cdots, 
L-1) \nonumber \\
\hat{\jmath}_{0} & = &  \tau_0\sigma_1 = \hat{\alpha}\sigma_1 \nonumber\\
\hat{\jmath}_{L} & = & \tau_L \sigma_{L+1} = \hat{\beta} \tau_L .
\end{eqnarray}
Here $\hat{\alpha} = \tau_0$ and $\hat{\beta} = \sigma_{L+1}$ represent a
set of independent random Boolean variables, which take the values 
$\{0,1\}$ with
expectations $\langle\hat{\alpha}\rangle = \alpha$ and
$\langle\hat{\beta}\rangle = \beta$, and which are drawn at every time step
from a uniform distribution. For later analysis it is convenient to transform
 the microdynamic equation (\ref{a1})-(\ref{a3}) to hole-occupation numbers 
$\sigma_i = 1 -\tau_i$, yielding
\begin{equation} \label{a2} 
\sigma_i' = \tau_i\sigma_{i+1} + \sigma_{i-1}\sigma_i
\qquad (i = 1,\cdots,L).
\end{equation}
The equations for the time evolution of averages $\langle\tau_i(t)\rangle
\equiv \langle\tau_i\rangle_t$, correlation functions
$\langle\tau_i\tau_k\rangle_t$ etc. can be derived by multiplying the
equations in (\ref{a1}) 
for different sites and subsequently averaging over arbitrary initial
configurations $\{\tau_i(0)\}$.
In this way one obtains an open hierarchy which couples the time changes of
a correlations function to higher order ones.

For large times ($t\rightarrow\infty$) the system will approach a
non-equilibrium stationary state, which is the main focus of attention
in this paper. Averages over the NESS are denoted by $\langle\cdots\rangle$.
This state is expected to be {\em unique}, i.e. independent of the initial
configuration $\{\tau_i(0)\}$. Therefore in analytical considerations
the initial state is always taken to
be the {\em empty} state $\{\tau_i(0)=0\}$ for all $i$ ($i = 1,2,\cdots,L)$,
which is also the most relevant initial state in physical applications, such
as traffic problems.

\subsection{Symmetries.}

A quantity of paramount interest is the average flux $<\hat{\jmath}_i>$ 
through the link ($i,i+1$), 
\begin{equation} \label{a4}
\langle\hat{\jmath_i}\rangle = \langle\tau_i\sigma_{i+1}\rangle,
\end{equation}
and the local flow velocity or average speed  $v_i =
\langle\hat{\jmath_i}\rangle/\langle\tau_i\rangle$. In the NESS these 
averages are independent of time, and the average flux can be calculated 
from the continuity equation, combined with (\ref{a3})  using the relation
$\langle\tau_i'\rangle=\langle\tau_i\rangle$. This yields a
constant site-independent flux through the system,
\begin{eqnarray} \label{a5}
<\hat{{\jmath}}_{i-1}> = <\hat{{\jmath}}_i>  = \langle\tau_i\sigma_{i+1}\rangle = j \nonumber \\
<\hat{{\jmath}}_0>  = \alpha(1-\langle\tau_1\rangle) = j
\nonumber \\
<\hat{{\jmath}}_L>  =  \beta\langle\tau_L\rangle = j
\end{eqnarray}
with $i=0,1,\cdots,L$. The flux in the NESS is translationally invariant.
Once the nearest neighbor correlations are known, the density profile can 
be calculated from
\begin{equation} \label{a4a}
\langle\tau_{i}\rangle = j + \langle\tau_i\tau_{i+1}\rangle.
\end{equation}
The equations of motion exhibit particle-hole symmetry, and the {\em
duality transformation},
\begin{eqnarray} \label{a6}
\tau_i & \leftrightarrow & \sigma_{L-i+1} \qquad (i = 1,\cdots,L)
 \nonumber \\
\tau_0 = \hat{\alpha} & \leftrightarrow & \sigma_{L+1} = \hat{\beta} \\
\hat{\jmath_i} & \leftrightarrow & \hat{\jmath}_{L-i} \nonumber,
\end{eqnarray}
maps the microdynamic equation (\ref{a1})-(\ref{a3}) into the equivalent 
representation
(\ref{a2}), i.e. the microdynamic equation is invariant
under particle-hole exchange. Consequently, the average occupation
numbers satisfy the symmetry relations
\begin{eqnarray} \label{a7}
\langle\tau_i\rangle (\alpha,\beta) & = & \langle\sigma_{L-i+1}\rangle
(\beta,\alpha) \nonumber \\
& = & 1 - \langle\tau_{L-i+1}\rangle (\beta,\alpha)
\end{eqnarray}
with $i=1,2,\cdots,L$. As the flux maps $\hat{\jmath}_i \leftrightarrow
\hat{\jmath}_{L-i}$ under the duality transformation, the average flux
satisfies the symmetry relation $j_i (\alpha,\beta) =
j_{L-i}(\beta,\alpha)$. However, the average flux in the NESS
 is constant for all sites, hence
\begin{equation} \label{a8}
j (\alpha,\beta) = j (\beta,\alpha).
\end{equation}
The particle-hole symmetry is a very powerful tool, as all properties for
$\alpha>\beta$ (high density) can be obtained from those for $\alpha<\beta$
(low density).

\section{Dynamics and Structures}
\label{dynamicssection}

\subsection{Build up of dynamic correlations.}

In this section we show that a qualitative analysis of the microdynamic
 equation and the resulting instantaneous configurations leads
already to the complete phase diagram, to an identification of the relevant
order parameters, and to a qualitative characterization of the structure of
the high and low density phase, as well as to the typical dynamics in the
different phases.

In the sequel we will consider the dynamics of {\it clusters} of particles
and holes described in terms of the Boolean variables,
\begin{eqnarray} \label{a9}
T_{kl} & = & \tau_k\tau_{k+1}\cdots\tau_l \nonumber \\
S_{kl} & = & \sigma_k\sigma_{k+1}\cdots\sigma_l
\end{eqnarray}
with $k<l$, i.e. a cluster has at least two constituents. The time evolution
equations for these objects are obtained by multiplying
(\ref{a1}) from $k$ to $l$, using $\sigma_j\tau_j = 0$.
The result is
\begin{eqnarray} \label{a10}
T_{kl}' & = & T_{k,l+1} + \tau_{k-1}\sigma_k T_{k+1,l+1} \nonumber \\
        & = & (\tau_k + \tau_{k-1}\sigma_k) T_{k+1,l+1} \\
S_{kl}' & = & S_{k-1,l} + S_{k-1,l-1} \tau_l\sigma_{l+1} \nonumber \\
        & = & S_{k-1,l-1} ( \sigma_l + \tau_l\sigma_{l+1}). 
\nonumber
\end{eqnarray}
Multiplication of both equations in (\ref{a10}) gives then
\begin{equation} \label{a11}
(T_{kn} S_{n+1,l})' = (\tau_k+\tau_{k-1}\sigma_k) T_{k+1,n+1} S_{n,l-1}
( \sigma_l + \tau_l\sigma_{l+1} ) = 0.
\end{equation}
where the relation $\tau_n\sigma_n = 0$ has been used.

The implications of (\ref{a11}) are quite interesting, as it
states that a configuration containing ($\cdots1100\cdots$) {\em cannot} be
created. As all possible configurations have evolved from the empty initial
state, configurations containing a cluster of particles {\it tailing} a 
cluster of holes  do not exist in the NESS.

Moreover, a configuration ($\cdots110100\cdots$) with a single
particle-hole pair separating the two clusters cannot be created either.
The reason is that only the nonexistent configuration
($\cdots?1100?\cdots$) in the previous time step could have created the
configuration under consideration. The question mark represents a "0" or a
"1".

Similarly a configuration ($\cdots11(01)^k00\cdots$) with $k$
($k=1,2,3,\dots$) intermediate hole-particle pairs does not exist, as it
could only have been created from the configuration
($\cdots?1(10)^k0?\cdots) = (\cdots?11(01)^{k-1}00?\cdots$). It then
follows by complete induction that none of the above configurations can
exist in the NESS.
 
Consequently, the possible configurations generated by the dynamics 
from the empty initial state do not contain any
configurations with two or more empty sites in front of the last cluster 
of particles. So, the fully parallel dynamics of the present TASEP 
builds up very {strong short range} correlations in the NESS.

From the observations about the build up of dynamic correlations, we 
arrive at some important conclusions about the structure of the NESS. Let
$k_0$ label the position of the tail particle in the last particle 
cluster\footnote{Note, however, that an instantaneous configuration of 
the whole system may not contain any cluster of particles (at low densities) 
or any clusters of holes (at high densities).}. The
configurations in the interval $[1, k_0-1]$ consist of {\it isolated} 
particles, separated by an arbitrary number of holes. In these so called {\em
free flow} configurations
\begin{equation} \label{a12a}
\tau_{i-1} \tau_i =0  \qquad (i < k_0),
\end{equation}
i.e. there is a `hard core repulsion' between particles on {\it nearest 
neighbor} sites.
The instantaneous fraction of occupied
sites (density) in this interval is therefore $\rho(<k_0)<1/2$.
The configurations in the interval $[k_0,L]$ consist of {\em isolated} holes,
separated by an arbitrary number of particles. In these so called {\em jammed} 
configurations
\begin{equation} \label{a12b}
\sigma_{i-1} \sigma_i =0  \qquad (i > k_0) ,
\end{equation}
i.e. there is a `hard core repulsion' between holes on nearest neighbor sites.
The instantaneous density in this interval $\rho(>k_0)>1/2$.

Therefore, {\em large} systems with an overall density (fraction of
occupied sites) $\rho<1/2$ necessarily have only bulk configurations of the
free flow type with a narrow {\it boundary layer} of jammed
configurations near the exit site of (yet unknown) width $\lambda_R<<L$, 
whereas for
$\rho>1/2$ bulk sites contain only jammed configurations with a narrow 
boundary layer of free flow configurations  of width $\lambda_L$ near 
the entrance site. Which
value of $\rho$ occurs, depends on the injection and removal rates,
$\alpha$ and $\beta$, and will be determined in section 4.1.

\subsection{Instantaneous profiles.}

To study the density profile we consider the dynamics of creation of pairs
and larger clusters. We first observe that the first particle pair can only
be created {\em  at} the exit site and only  if $\beta<1$. This
follows from the evolution equation (\ref{a10}) for $T_{k,k+1}$
which shows that creation of a new pair at ($k,k+1$) requires the existence
of a pair at ($k+1,k+2$). Therefore, creation of the {\em first} pair $T_{L-1,L}$
at $t+1$ is governed by
\begin{equation} \label{a12}
(\tau_{L-1}\tau_L)' = (\tau_{L-1} + \tau_{L-2}\sigma_{L-1})\, \tau_L
\, (1-\hat{\beta}),
\end{equation}
where the term $\tau_{L-1}(t)\tau_L(t)$ on the left hand side of the 
equation vanishes, as this pair has not yet been created.
This implies that pairs cannot be created if $\langle\hat{\beta}\rangle =
\beta =1$. Consequently, there is no boundary layer near the exit,
the density profile is totally {\em flat} over the whole system and 
all configurations are
pure free flow configurations. All space and time dependent correlations in
this NESS can be calculated exactly in a simple manner, as will be shown in
section 5.2.

Next we consider the case $\beta<1$, where particle clusters can be
created, at least near the exit site. In case the injection rate is
smaller than the removal rate, 
 the average interval between arrivals $1/\alpha$ at the pile up region
 near the exit 
is larger than the average interval $1/\beta$ between removals, and
a large fraction of configurations are pure free flow configurations without
any clusters near the exit site. On average there is only a narrow
 boundary layer of jammed configurations. So, the bulk properties of
the system in the low density case $\alpha<\beta$ are determined by the
injection rate $\alpha$ at the entrance site.

In the jammed phase ($\alpha>\beta$ ), there is on average a large
backup starting near the exit. The jammed configurations in the NESS
cover the bulk of the system, leaving only a narrow boundary
layer with free flow configurations near the entrance. The bulk 
properties in the high density phase are determined by
the removal rate $\beta$ at the exit site.

In case $\alpha=\beta$, there exist downstream free flow configurations 
and upstream jammed configurations, which occupy on average a finite fraction of the
system. There are in fact two coexisting phases, spatially separated by an 
`external driving field', imposed by the reservoirs with injection and 
removal rates $\alpha=\beta$.
The phases are separated by an interface of microscopic size, containing
 only configurations with alternating particles and holes. These 
 configurations, when present in a finite fraction of the system, would form 
the so called `maximal current phase' \cite{derrida}. In the present model
the width of  this 
interface diverges, i.e. attains a macroscopic width, only at a single
point in the phase diagram, where $\alpha =\beta \rightarrow 1$ (see section 4.2).                                                                                                                                                                                                                                  In section 4.2 we will return to the dynamics and the structure of the
interface.
In the phase diagram of the stochastic model of \cite{rsss} there is a line $(1-\alpha ) (1-\beta) = 1-p$, where the density profile is {\it flat} over the whole system.
In the present deterministic model $(p=1)$ this corresponds to the
line $\beta = 1$ (free flow phase), and to the line $\alpha = 1$
(jammed phase).

\section{Phase Diagram}

\subsection{Free flow and jammed phases}

For large system size $L$ (thermodynamic limit)  and $\alpha<\beta$
the system is in the free
flow phase , and  the dynamics rigorously implies that
the bulk of the system (except the 
boundary layer near the exit) has only {\em isolated} particles. So, 
this low density phase is characterized by an
average density $\rho_F<1/2$ and has the
{\it vanishing} order parameters $\langle\tau_i\tau_{i+1}\rangle =
\langle\tau_i\tau_{i+1}\tau_{i+2}\rangle = \cdots = 0$ for $i$ in the 
{\em bulk} phase, defined as $i<<L-\lambda_R$
with $\lambda_R$  the width of the boundary layer  near the exit.
In fact, even the microscopic order parameter $\tau_i\tau_{i+1} = 0$
in the free flow phase on the basis of (\ref{a12a}). Of course the 
 correlations $\langle\sigma_i\sigma_{i+1}\rangle$,
$\langle\sigma_i\sigma_{i+1}\sigma_{i+2}\rangle$ etc. are {\em
non-vanishing} in this phase.

Moreover, the vanishing order parameter $\langle\tau_i\tau_{i+1}\rangle =
0$ for bulk sites, in combination with  relations 
(\ref{a5}) imply the following properties of the low density
{\em free flow} phase:
\begin{eqnarray} \label{a13}
j = \langle\tau_1\rangle = \langle\tau_2\rangle = \cdots =
\langle\tau_i\rangle = \rho \nonumber \\
\langle\tau_1\rangle = j = \rho = {\alpha}/(1+\alpha) \nonumber \\
\langle\tau_L\rangle = {j}/{\beta} = {\alpha}/[\beta(1+\alpha)],
\end{eqnarray}
where $i$ is a bulk site.
Several interesting features can be seen. First of all, the bulk dynamics
is completely determined by the input rate $\alpha$ at the entrance site, 
as already explained in
section 3.2. Secondly, the flux  equals the bulk 
density, indicating that particles are never blocked
in the free flow phase, and are traveling with an average speed 
$v_F \equiv j/<\tau_i> =1 $. A particle entering the lattice is 
never blocked until it leaves the
bulk and enters the  boundary layer near the exit, where it slows down to a
velocity $v_J = j/\langle\tau_L\rangle = \beta$ with $\beta < 1$ as a 
consequence of the pile up.
The bulk density of the system, $\rho=\alpha/(1+\alpha)$, is always smaller
than 1/2, since $\alpha<\beta \leq 1$. Therefore, we will also refer to the
phase with $\alpha<\beta$ as the {\em low density phase} of the system.

Similarly there is a {\em jammed} phase for $\alpha>\beta$ of high
density $\rho>1/2$, containing only isolated holes except in a
boundary layer of width $\lambda_L$ near the entrance. 
The phase is characterized by the
microscopic order parameter $\sigma_i\sigma_{i+1}=0$, or equivalently by the
non-vanishing order parameters $\langle\tau_i\tau_{i+1}\rangle$,
$\langle\tau_i\tau_{i+1}\tau_{i+2}\rangle$, etc. in the bulk.

The properties of the high density phase with $\alpha>\beta$ can be 
related to those of the low density phase with $\alpha < \beta$ by 
the particle-hole symmetry of section 2.2. In this case, particle-hole 
symmetry implies that the dynamics of particles moving forward is
 identical to that of the holes moving backwards. In this point of 
view holes are injected at the exit site with probability $\beta$ 
and move downstream where they are finally removed from the lattice 
with probability $\alpha$. So, (\ref{a13}) implies
for the {\it jammed} phase:
\begin{eqnarray} \label{a14}
\rho & = & \langle\tau_L\rangle (\alpha,\beta) = \langle\sigma_1\rangle
           (\beta,\alpha) \nonumber \\
     & = & 1 - \langle\tau_1\rangle (\beta,\alpha) = 1/(1+\beta) \\
\langle\tau_L\rangle & = & \langle\tau_{L-1}\rangle = \cdots =
                       \langle\tau_i\rangle = \rho \nonumber \\
\langle\tau_1\rangle (\alpha,\beta) & = & 1 - \langle\tau_L\rangle (\beta,
\alpha) =  1 - \beta/[\alpha(1+\beta)] \nonumber,
\end{eqnarray}
where $i$ is a bulk site in the jammed phase, defined as $ i >> \lambda_L$.
Similarly the average flux in the jammed phase follows from
(\ref{a8}) as
\begin{equation} \label{a15}
j = j_J (\alpha,\beta) = j_F (\beta,\alpha) = \frac{\beta}{1+\beta} = \beta
\rho = 1- \rho,
\end{equation}
which implies an average speed $v_J \equiv j/\rho = \beta$. In the boundary
layer near the entrance a particle has a higher average speed, $v_1 = j/\langle\tau_1\rangle$, and it slows down to speed 
$\beta$, because of frequent blockage by a preceding particle. 
The average flux may also
be written as $j = 1-\rho$, indicating that the flux in the jammed phase
equals the density of holes, which move with unit speed in backward 
direction. We note 
that these results are exact in the thermodynamic limit as $L \rightarrow 
\infty$.

In summary, the phase diagram, where $ 0\leq \{ \alpha, \beta\} \leq 1$, has a high density phase ($\rho>1/2$) for $\alpha > \beta$, and a low density phase ($\rho < 1/2$) for $\alpha < \beta$. On the line $\alpha=\beta$ there are two coexisting phases, separated by an interface with an {\it instantaneous}
position $R$: a low density phase can be found downstream of the interface 
and a high density phase upstream,
each occupying a finite fraction of the system (as already discussed in
section 3) with  a jump $\bigtriangleup\rho$ in the
density  between the two regions. In open systems
 with fixed $\alpha$ and $ \beta$ the position $R$ wildly fluctuates, as it
 can be anywhere on the lattice with uniform probability. For a system of 
$L= 1000$ sites this statement has been verified by collecting $1.2 \times 10^7$ measured values of $R$ into 10 equal-sized bins. The resulting histogram is flat within fluctuations of $1\%$.

Let us  compare the
behavior of flux $j$ and bulk density $\rho$ when crossing the transition
line $\alpha=\beta$, where $\rho=\alpha/(1+\alpha)$ and $j=\rho$ in the low density phase, and $\rho=1/(1+\beta)$ and $j=1-\rho$ in the high density phase. This shows that the flux $j$ is continuous
across the line $\alpha=\beta$, whereas the density makes a jump
$\bigtriangleup\rho = \rho_J - \rho_F = (1-\alpha)/(1+\alpha)$. Therefore the
NESS of this model shows a first order phase transition across the line
$\alpha=\beta$. Figure 1 shows the flux $j(\rho)$, as a 
function of the average density at a fixed value of the removal rate $\beta$.
It is the analog
of the equation of state for the pressure $p(\rho)$ in thermal equilibrium.
As $\alpha$ increases to $\beta$, the flux $j(\rho) =\rho$, as well as the 
density increase up till $\rho_F = \beta /(1+\beta)$. In the coexisting phase 
region ($\alpha = \beta$) the flux remains constant, and the bulk density 
is given by
\begin{equation} \label{a16}
\rho = \frac{1}{L} \sum_i <\tau_i> = \left( \frac{R}{L} \right) 
\frac{\alpha}{1+\alpha} + \left( 1 - \frac{R}{L} \right) \frac{1}{1+\alpha}.
\end{equation}
It may be anywhere between $\rho_F$ and $\rho_J = 1/( 1+ \beta)$, as the location 
$R$  hops around the interval $(1,L)$.
As $\alpha$ increases further $(\beta < \alpha < 1)$, the density increases 
from $\rho_J$ to 1, and the flux $j(\rho) = 1 -\rho $ decreases. The triangle
 bounding the region $ (j < {\rm min} \{ \rho, 1-\rho\})$ is called the {\it 
fundamental diagram} in traffic problems, and corresponds to the coexistence
region in thermodynamic phase transitions. As the interface position $R$ in
the open system, measured over a very long period of time,
 is uniformly distributed over all sites of the system, one obtains in the coexisting phase region a linear
density profile $\rho (x)$ by averaging the instantaneous profiles
$\hat{\rho} (x | R)$ over all $R$, yielding
\begin{equation}
\rho (x) = \frac{1}{L} \sum^{L}_{R=1} \hat{\rho} (x|R)= \rho_F
(1- \frac{x}{L}) + \rho_J \frac{x}{L}. 
\end{equation}
A more detailed discussion of the dynamics of the
interface will be given in the next section. 

Before concluding this section we discuss the collective dynamics in the
different bulk phases. In the low density or free flow phase every
particle is preceded by at least one hole. Consequently it will advance one
site per time step, and has the maximum speed
$v_{F} = 1$. In fact any free flow configuration with a hole in front of it,
 is propagated as a whole with velocity $v_{F} = 1$.
In the jammed phase on the other hand, every cluster of $c$ particles is
preceded by a single hole. At every time step {\em only} the lead particle
of each cluster advances one site, and becomes the tail particle
of the preceding cluster. Therefore clusters of constant length are moving
backwards with unit speed, and so do isolated holes.
The $c$-th particle in a cluster
makes its first move only after $c$ time steps, and the average speed of a particle $v_J= \beta$ (see (\ref{a15})). The motion in the jammed
phase is therefore characterized by {\em start-stop waves}, which are
typical for most congested traffic flows.

It is interesting to compare the results (\ref{a13})--(\ref{a15}) for fully 
parallel updating with those for {\it sublattice--parallel} updating, 
as derived
in (A5)-(A7) of the Appendix. Let $<\tau_+>$ and $<j_+>$
be respectively  the bulk density and the flux at even sites, as defined 
in the Appendix,
and $<j_->$ and $<\tau_->$ the corresponding ones for  odd sites, then we 
have for
the sublattice--parallel dynamics in the {\it free flow} phase 
$(\alpha < \beta)$,
\begin{eqnarray} \label{a16a}
<\tau_+> & = & \alpha \qquad <\tau_-> = 0  \qquad <\tau_L>  =\alpha/\beta 
\\ \nonumber
 <j_+> & = &\alpha \qquad <j_->=0,
\end{eqnarray}
and for the {\it jammed } phase $(\alpha > \beta)$
\begin{eqnarray} \label{a16b}
<\tau_+> &=& 1 \qquad <\tau_-> = 1- \beta  \qquad <\tau_1>  =
1 -\beta / \alpha
\\ \nonumber
<j_+> & =& \beta  \qquad <j_->=0.
\end{eqnarray}
The average properties, in particular of the odd sites, are very different from the
TASEP with fully synchronous dynamics.

\subsection{Coexisting phases}

In this section we  study the dynamics of the interface separating the
 low and high
density phases in the coexisting phase region. To define the instantaneous
location $R$ and the instantaneous width $w$ of the interface we
introduce its front and tail sites $l_0$ and $j_0$. Let $j_0$ be 
the position of the front hole in the first/most advanced cluster of holes, 
and $l_0$ that of the tail particle in the last/least advanced cluster of 
particles, where  $j_0 < l_0$ (see section (3.1)).
The sites $j_0$ and $\ell_0$ belong with certainty to the 
low and high density phase
respectively. Then the interval ($j_0,\ell_0 = j_0+2n+1$) contains only $n$
alternating particle-hole pairs $(10)^n$ ($n=0,1,2,\cdots$), which are
allowed in both phases. The instantaneous position of the interface is
defined as $R = \frac{1}{2}(j_0+l_0)$ and its instantaneous {\em width} 
as $w(n) = 2n = l_0-j_0-1$. As the particle (c.q. hole) cluster is
moving with unit speed forward (c.q. backward), the width decreases
by 2 units per time step, vanishing after $n$ times steps and yielding a
hole-cluster adjacent to the particle-cluster. During this period
$R=\frac{1}{2}(j_0+\ell_0)$ remains fixed.

What happens next depends on the sizes $h$ and $c$ of the two adjacent
hole- and particle clusters respectively. If $h> c$, the positions $j_0$ 
and $l_0$ remain
fixed during ($h-2$) time steps. After ($h-1$) time steps the cluster of
particles disappears, and $j_0$ moves one site forward, whereas
$l_0$ makes a forward {\em jump} to the last particle on the then 
last  cluster. If
$h < c$, the same statements can be make with $l_0$/$c$/particles/forward
interchanged with $j_0$/$h$/holes/backwards respectively. If $c=h$ both points
$j_0$ and $l_0$ {\em jump simultaneously}. The dynamics of the interface 
width is illustrated in Figure~2.

The probability distribution for the sizes of the right and left jumps, as
well as those for the time intervals between the jumps, are determined by
the probability distribution of finding $n$ particle-hole pairs between two
particle clusters in the high density region, or between two hole clusters 
in the low density region.
The position $R$ of the interface performs a random walk over all sites
of the system around the average position $\langle R\rangle=L/2$. Once
the above probability has been calculated, its mean square displacement
$(\delta R)^2 = \langle(R-L/2)^2\rangle$ and the associated short time
diffusion
coefficient $\cal{D}$ can be calculated in principle for time intervals $T$
 satisfying the inequality $\delta R = \sqrt{2{\cal D}T}
< <L/2$. The long time diffusion coefficients vanishes due to the presence 
of the boundaries. Here we only illustrate the basic idea of the method by
calculating the average width $\langle w\rangle$ of the interface, using 
a crude mean field estimate.

The width $\langle w\rangle$, measured over a long time interval, is shown
in Figure 3, as a function of the injection rate.
This behavior can be understood on the basis of simple mean field
arguments. Let the
instantaneous interface configuration be ($\cdots?00(10)^n11?\cdots)$ with
($n=0,1,2,\cdots$), then its width is $w(n) = 2n$. The probability on the
configuration $(00(10)^n)$ tailing 
$(11?\cdots)$ -- which is the start of the jammed phase -- is $P(n) = 
(1-\alpha)\alpha^n$, where $\alpha^n$ is the probability
for injecting $n$ particles and ($1-\alpha$) the probability for not
injecting a particle. In the present asymmetric exclusion process, every
injected particle is followed by a hole. The average interface width is
then
\begin{equation} \label{a17}
\langle w\rangle = \sum_n {2nP(n)} = 2/(1-\alpha).
\end{equation}
This estimate gives a fair estimate of  the simulation results, as shown in Figure 3.

\section{Correlation Functions}
\subsection{Profiles and nearest neighbor correlations}

Consider first the low density or free flow phase where $\alpha < \beta$. To 
determine the density profile $<\tau_k >$ from (\ref{a4a}) we need the nearest
neighbor correlation function $<\tau_k \tau_{k+1}>$, which represents the
probability of finding the sites $(i,i+1)$ occupied.
As already derived in section 4.1, we have $<\tau_k > = \rho$ and $<\tau_k \tau_{k+1}> = 0$ for {\it bulk} sites
$(k \ll L - {\lambda_R} )$.  It remains to
calculate these quantities in the  boundary layer 
$k \stackrel{>}{\sim} L - \lambda_R$ near the exit.
To do so, we need the dynamics (\ref{a10}) of the cluster correlation functions
(\ref{a9}), averaged over the NESS, where $< T^\prime > = <T>$. This yields
\begin{equation}\label{d1}
<T_{k\ell}> = <T_{k,\ell +1}> + <\tau_{k-1} \sigma_{k} T_{k+1,\ell +1}> ,
\end{equation}
where $\ell = k+1,\ldots ,L$ and we recall that $\tau_{L+1} = 1- \hat{\beta}$.
If $T_{k+1,\ell +1}$ refers to the least advanced particle cluster, then
the sites $(k-1,k)$ belong by definition to the boundary layer, and 
the probability for the
configuration $(\tau\sigma T)_{k-1,\ell +1}$ equals $\alpha$ times the
probability for the configuration $T_{k+1,\ell +1}$. Since $k+1$ is by
definition the last site on the least advanced particle cluster, 
the occupation number $\sigma_k $ equals unity 
with probability 1. In fact there are only very few particle clusters 
in the pile up region near the exit site, as the average removal interval $1/\beta$ is less than
the average arrival interval $1/\alpha$ at the pile up region. So, we expect 
that the least advanced particle cluster gives the dominant contribution to 
the probabilities, and we make the {\it mean field assumption} that the above 
factorization holds for all further advanced particle clusters
as well, i.e.
\begin{equation}\label{d2}
<T_{kl}> = <T_{k,l +1}> + \alpha <T_{k+1,l+1}>.
\end{equation}
The present mean field approximation therefore assumes 
that 4-point correlations and higher order ones between a 
particle-hole pair 
and the particle cluster just in front of it 
(which by definition belong to the boundary layer 
in the present model) are negligible in the low density phase.
 
The recursion relation above can be solved starting from $l = L$, where
$<T_{k,L+1}> = $   $<~T_{kL}> (1-\beta )$, and yields after iteration
\begin{equation}\label{d3}
<T_{kL}> =  \frac{\alpha (1-\beta )}{\beta} <T_{k+1,L}> \nonumber \\
     =  \left( \frac{\alpha (1-\beta )}{\beta}\right)^{L-k} <\tau_L>,
\end{equation}
where we have used the relation $<T_{L,L}> = <\tau_L>.$ Taking $\ell = L-1,
L-2$, etc. gives $<T_{k,L-1}> = (\alpha / \beta ) <T_{k+1,2}>$, etc. and one
finds by complete induction
\begin{equation}\label{d4}
<T_{k,L-\ell} > = \left( \frac{\alpha}{\beta}\right)^l <T_{k+l,L}> .
\end{equation}
Combining this with (\ref{d3}) yields
\begin{equation}
\label{d5}
T_{k\ell} = <\tau_k \tau_{k+1} \ldots \tau_{l}> 
= \left(\frac{\alpha}{\beta}\right)^{L-k} (1-\beta )^{l-k}
< \tau_L>.
\end{equation}
The density profile is then obtained by inserting (\ref{d5}) for $l = k+1$ 
in (\ref{a4a}) with the result
\begin{equation} \label{d6}
<\tau_k > = j \{ 1 + \frac{1-\beta}{\beta} \left( 
\frac{\alpha}{\beta}\right)^{L-k} \},
\end{equation}
where $<\tau_L>=j/\beta = \alpha / [\beta (1+\alpha)]$ has been used. 
By taking the thermodynamic limit of (\ref{d5}) and (\ref{d6}) one recovers
the properties of section 3 and 4 for the bulk phase, i.e. the order
parameters $< \tau_k \tau_{k+1}> = <\tau_k \tau_{k+1} \tau_{k+2}, \ldots>$
are vanishing, and the bulk density is $\rho = < \tau_k > = j$. In
addition (\ref{d6}) describes the profile of the {\it  boundary 
layer} near the exit.
The excess density, which equals the nearest neighbor correlation 
$< \tau_k \tau_{k+1} >$, decreases exponentially on a length scale
 1/ ln $(\beta / \alpha )$, {\it independent} of the system
size. 

Our mean field assumption, formulated below (\ref{d1}), also implies that site $k$ in 
$T_{kl}$ marks the beginning  of the boundary layer. Hence, the 
probability $P(k)$ that the width of the boundary layer is $L-k$, is 
then {\it proportional} to the excess density $<\tau_i> -j$ in (\ref{d6}), 
yielding after proper normalization,
\begin{equation}
\label{d61}
P(k) = \zeta^{L-k} / (1-\zeta)
\end{equation}
with
\begin{equation}
\label{d62}
\zeta = \alpha/\beta.
\end{equation}
The expected width of the boundary layer in the low density phase is then,
\begin{equation}
\label{d63}
\lambda_B = <L-k> = \frac{\zeta}{1-\zeta} = \frac{\alpha}{\beta - \alpha},
\end{equation}
and the fluctuation $\delta \lambda_B$ around this average is 
\begin{equation}
\label{d64}
(\delta \lambda_B)^2 = <(L-k)^2> - <L-k>^2 = \frac{\zeta}{(1-\zeta)^2}
=\frac{\alpha \beta}{(\beta - \alpha)^2}.
\end{equation}
The width of the boundary layer is `microscopic' in nature, i.e. 
independent of the system size. In the limiting case as $\alpha 
\uparrow \beta$ (coexisting phases) the width $\lambda_B$ diverges 
and becomes of macroscopic size. It is given by $L-R$, as the location 
$R$ hops around  over all sites of the lattice.

The correlation functions in the high density or jammed phase with
$\alpha > \beta$ can be obtained from  particle-hole symmetry, e.g.
$<\tau_k \tau_{k+1}> (\alpha , \beta ) = < \sigma_{L-k} \sigma_{L-k+1}> 
(\beta , \alpha )$
and yields
\begin{equation}\label{d7}
< \tau_k \tau_{k+1}> = \frac{1-\beta}{1+\beta} - \frac{1-\alpha}{1+\beta}
\left( \frac{\beta}{\alpha}\right)^i.
\end{equation}
Similarly, the density profile follows from (\ref{d7}) and (\ref{a4a}) as
\begin{equation}\label{d8}
< \tau_k > = \frac{1}{1+\beta} \left[ 1- (1-\alpha ) 
\left( \frac{\beta}{\alpha} \right)^k \right].
\end{equation}
It shows a boundary layer near the entrance site with a deficit 
density that decays 
on a length scale $1/\ln ( \alpha / \beta )$.

For the coexisting phase region, where $\alpha = \beta$, the boundary layers 
at both ends diverge, i.e. become of macroscopic size, and the typical
width $\xi$ of the interface $\xi = <w> = 2 / (1-\alpha )$, or
equivalently $ \sim 1/ | \ln \alpha |$, as was calculated in section 4 on the
basis of dynamical considerations.

In Figure 4 the theoretical profiles  for 2- and 3-point correlations are compared
with those measured in computer simulations, 
for two different combinations of parameter sets $\{ \alpha = 0.50; 
\beta = 1 \}$ and $\{ \alpha = 0.50; \beta = 0.51\}$
and the results are {\it indistinguishable}. For
 $\beta = 1$ the profile is flat, as already explained below (\ref{a12}).
Moreover we have compared simulations and theoretical predictions for the 
three- and
four-point correlation functions $< \tau_k \tau_{k+1} \tau_{k+2} \ldots>$
for $\alpha = 0.50$ and $\beta = 0.55$. The
former ones are shown in Figure 4b, and the results are again 
indistinguishable.

In the Appendix  the density profile and correlation functions for the same 
TASEP  with sublattice--parallel updating \cite{schuetz} have been calculated, 
using the same mean field theory.
Here the average flux is quite different, i.e. $j = \beta <\tau_L > = \alpha$,
but the profile $<\tau_k>/j$ is quite similar, i.e.
\begin{equation}\label{d9}
< \tau_k > = j \left\{ \delta_{k,{\rm even}} + \frac{1-\beta}{\beta}
\left( \frac{\alpha}{\beta}\right)^{L-k} \right\},
\end{equation}
at least for even sites.
For comparison we also  quote the results for the correlation functions,
as derived in (A11) of the appendix, i.e.,
\begin{eqnarray} \label{d10}
<T_{kl}> & = & < T_{k,l -1} > = \frac{\alpha}{\beta} < T_{k+1,l} > =
\frac{\alpha}{\beta} < T_{k+1,l -1} > \nonumber \\
& = & \left( \frac{\alpha}{\beta} \right)^{L-k} 
(1-\beta )^{\frac{1}{2}(l-k)} <\tau_L>,
\end{eqnarray}
which should be compared with (\ref{d5}).
We emphasize that $k,l$ and $L$ are {\it even} numbers in the present formulas.
Comparison with (\ref{d5}) shows that the correlation functions are quite different,
in particular the differences between {\it even} and {\it odd} sites.
The results (\ref{d9}) and (\ref{d10}) are {\it exact} to terms of
order $(\alpha / \beta )^L$, as can be verified by
comparison with the exact results derived in Refs. \cite{schuetz,hinrichsen}.

We {\it conjecture} that the corresponding results 
(\ref{d3}) - (\ref{d8}) for 
the same TASEP with fully parallel updating and
with open boundaries are exact as well, up to exponential terms of order
$( \alpha / \beta )^{L}$ for $ \alpha < \beta$, and up to
$(\beta / \alpha )^{L}$ for $\alpha > \beta$.
However, for $\alpha$ close to $\beta$, say $\alpha =\beta ( 1- \delta/L)$, 
the expressions for the profiles in the boundary layers break down, as the neglected terms  $(\alpha/ \beta)^L \simeq \exp(-\delta)$ become of 
${\cal O}(1)$ for large $L$.

\subsection{Spatial and Temporal Correlations}

Next we will determine the correlation functions $<\tau_i \tau_{i + R}>$ 
for {\it bulk} sites in the low-density phase $( \alpha < \beta \leq 1)$.
 As explained in section 4.1 the dynamics rigorously implies that the 
`microscopic' order parameter $\tau_i \tau_{i+1} = 0$ in the bulk phase. 
Consequently, the microdynamic equation  for all sites outside 
the pile up region near the exit becomes
\begin{eqnarray} \label{d11} 
\tau^{\prime}_{1} & = & \tau_0 \sigma_1 = \hat{\alpha} (1-\tau_1) \nonumber \\
\tau^{\prime}_{i} & = & \tau_{i-1} \qquad (1< i << L - \lambda_R).
\end{eqnarray}
The last relation implies {\it translational invariance} of the
correlation for the bulk phase, i.e.
\begin{eqnarray}\label{d12}
<\tau_i \tau_{i+R}> & = & < \tau_{i-1} \tau_{i+R-1} > = \ldots = \nonumber \\
& = & < \tau_1 \tau_{1+R}> = \alpha <\tau_R> - \alpha < \tau_1 \tau_R >.
\end{eqnarray}
The last equality follows from (\ref{d11}) for $\tau_1$.
Then (\ref{d12}) implies that the pair distribution function, defined as $g(R)
=< \tau_1 \tau_{1+R}> / < \tau_{1+R}>$, satisfies the recursion relation
\begin{equation}\label{d13}
g(R+1) = \alpha - \alpha g(R).
\end{equation}
It follows straightforwardly from (\ref{d13}) that the generating
 function has the form:
\begin{equation} \label{d14}
F(z) \equiv \sum^{\infty}_{R=0} z^R g(R) = \frac{1-z(1-\alpha )}{(1+\alpha z)(
1-z)}.
\end{equation}
By extracting the coefficient of $z^R$ from (\ref{d14}) we find the pair
correlation function in the {\it bulk} of the low density phase,
\begin{eqnarray}\label{d15}
< \tau_i \tau_{i+R} > & \equiv & < \tau_i > <\tau_{i+R}> [1+C(R)] \nonumber \\
& = & < \tau_1 >^2 [1- (- \alpha )^{R-1} ].
\end{eqnarray}
This is an asymptotically exact result for 
$1 \leq \{ i,i+R \} \ll L-\lambda_R$.  In case $\beta = 1$
there is no right boundary layer, and (\ref{d15}) is exact for all interparticle
distances $R$.
The pair function oscillates around its uncorrelated value, it vanishes for
$R=1$, as it should, and is independent of $\beta$, because the bulk phase is
only determined by the injection rate $\alpha$. Neighbor particles are strongly correlated.
The spatial correlation function in the bulk $C(R) = - (-\alpha )^{R-1}$
decays exponentially with a correlation length $\xi_F = 1/ |\ln \alpha |.$
This correlation length becomes of macroscopic size as $\alpha \rightarrow 1$ 
and $\beta = 1$, where the system approaches the critical maximal current 
phase \cite{derrida}, which is in the present model only realized at the parameter values $\alpha = \beta = 1.$

The result (\ref{d15}) for the bulk correlations is very different from those
in Refs. \cite{schuetz,hinrichsen} for sublattice--parallel dynamics,
 where the spatial correlations 
do not depend on the interparticle distance $R$. For instance, in that model
 one has in the {\it bulk} of the low density phase
\begin{equation}\label{d16}
< \tau_i \tau_{i+R} > = \left\{ \begin{array}{ll}
< \tau_1 >^2 (1+\alpha )  & \mbox{($i,R$ even)} \\
                   \qquad 0 & \mbox{(elsewhere)}.
\end{array} \right.
\end{equation}
So far we have discussed spatial correlation functions for bulk 
sites of the NESS. Similar results can be obtained for
{\it time} and {\it space}-dependent correlation functions. It follows from
the microdynamic equation (\ref{d11}) that  in the
 low density phase
\begin{equation} \label{d17}
\tau_{i+R}(t) = \tau_{i+R-1} (t-1) = \ldots = \tau_{i+R-t} (0),
\end{equation}
and consequently we have the asymptotically exact result,
\begin{eqnarray}\label{d18}
< \tau_{i+R}(t) \tau_i (0) > & = & < \tau_{i+R-t} \tau_i > \nonumber \\
& = & < \tau_1 >^2 \left\{ 1-(-\alpha )^{R-t-1} \right\},
\end{eqnarray}
valid for $ 0 \leq \{ i,j \} \ll L - \lambda_R$ with $j=i + R-t $ and
 $R \neq t$. For the special case
$\beta = 1$ the result (\ref{d18}) is exact for all sites and for all times.

One may also consider correlation functions involving fluxes. In the
{\it low density} phase $(\alpha < \beta )$ we have
\begin{equation}\label{d19}
< \tau_{i+R} (t) \tau_i (0) > = < \hat{\jmath}_{i+R}(t) \tau_i(0)> = 
< \hat{\jmath}_{i+R}(t) \hat{\jmath}_i (0) >,
\end{equation}
because the macroscopic order parameter $\tau_i \tau_{i+1} =0$.
The result obtained for the low density phase can be extended easily to the
{\it high density} phase using particle-hole symmetry and the relation 
$\hat{\jmath} =\sigma_{i+1} $ on account of (\ref{a2}) and (\ref{a12b}).
This yields for bulk sites
\begin{eqnarray}\label{d20}
&< \tau_i \tau_{i+R}> (\alpha ,\beta )  &=  < 1- \tau_{L-1+1} -
\tau_{L-R-i+1} - \tau_{L-i+1} \tau_{L-R-i+1} > (\beta, \alpha ) \nonumber \\
& &=  \frac{1}{(1+\beta )^2} \{ 1-(-\beta )^{R+1}\} = < \tau_i >^2
\{ 1- (-\beta )^{R+1} \} \nonumber \\
& < \tau_i \hat{\jmath}_{i+R} > (\alpha , \beta )  &= < \tau_i \sigma_{i+R+1} > (\alpha ,
\beta ) \nonumber \\
& & =  < \tau_{L-R-i} (1-\tau_{L-i+1}) > (\beta , \alpha ) \nonumber \\
& & = j < \tau_i >   \{ 1- (-\beta )^{R+1} \} \nonumber \\
&< \hat{\jmath}_i \hat{\jmath}_{i+R} > (\alpha , \beta ) &= 
<\sigma_{i+1} \sigma_{i+R+1}> = j^2 \{ 1-(-\beta )^{R-1} \}.
\end{eqnarray}
For the {\it coexistence} region $(\alpha = \beta )$ extensive
measurements were made. It appears that the properties such as density, flux, 
etc. for both regions are equal to the properties in their 
associated phases. It seems interesting to measure the correlations 
between site $i$ in the low density region and site $i+R$ in the high 
density region. However, due to the random nature of the interface
location, the occupation numbers $\tau_i$ and $\tau_{i+R}$ appear to be
uncorrelated, even if $R \leq 1/ |\ln \alpha |$.

\subsection{Profile on coexistence line}

In section 3 the profile and interface between low and high density phase
$(\alpha = \beta )$ has been studied on the basis of purely dynamical
considerations. A few  additional results can be obtained from particle hole
symmetry (\ref{a6}) and the constant flux relation (\ref{a5}).

On the transition line $\alpha = \beta$, it follows from (\ref{a7}) that
$<\tau_i> = 1 -$ $<\tau_{L-i+1}>$. The density in the middle of
the lattice is therefore on average 1/2. The average number of 
particles $<N>$ on the
lattice follows by summing the above relation over all sites, and yields
\begin{equation}\label{d21}
<N> = \sum_{i=1}^{L} <\tau_{i}> = \textstyle{\frac{1}{2}}L.
\end{equation}
The flux $j = \alpha / (1+\alpha )$, which is  continuous across the transition
line (see section 3.1), allows us to calculate the boundary values exactly
\begin{equation}\label{d22}
\begin{array}{lcl}
<\tau_1> = {\alpha}/{(1+\alpha)} \quad & ; & \quad <\tau_L> = {1}/{(1+\alpha)}
\nonumber \\
<\tau_1 \tau_2> = 0 \quad & ; & \quad <\sigma_L \sigma_{L-1}> = 0.
\end{array}
\end{equation}
There are two coexisting phases, downstream of the interface the low density
phase with $\rho_F = \alpha / (1+\alpha )$ and upstream the high density phase
with $\rho_J = 1/(1+\alpha )$, separated by an interface located at $R$. 
This interface hops around over the whole lattice, where two 
instantaneous profiles (at $t_1$ and $t_2$) are  shown in Figure 5 
for $\alpha = \beta = 0.50$.

\subsection{Travel times}

This section studies the travel time of the particles, which is by definition
the number of time steps that have passed from the moment a particle enters
the lattice until it leaves the lattice. Because particles cannot travel with
a speed larger than unity, the actual travel time will always be larger
or equal to the size of the lattice $T \geq L$. For the average travel time, 
the following definition is used:

\begin{equation}\label{d23}
<T> = \sum^{L}_{i=1} {<\tau_i >}/{j} = {<N>}/{j},
\end{equation}
where $<N>$ denotes the average number of particles on the lattice. The
resulting expression implies that the flux is equal to the average (unknown)
number of particles $<N>$ divided by the average travel time $<T>$.

The present sections will give analytical and simulation results for
travel times in low- and high density phases, as well as 
travel times in the coexistence region.
In the {\it low-density phase} of a large system $(L \rightarrow
\infty )$, the speed in the bulk equals unity. Therefore we expect the
average travel time to be approximately equal to the size of the lattice,
$<T> \simeq L$. It can be calculated directly from (\ref{d23}) and
the density profile in (\ref{d6})
 and yields in the low density phase $(\alpha < \beta )$,
\begin{eqnarray} \label{d24}
<T> &=& L + \frac{1-\beta}{\beta -\alpha} \left[ 1 - \left( 
\frac{\alpha}{\beta}
\right)^L \right] \nonumber \\
& = & L + \frac{1-\beta}{\beta - \alpha} \qquad (L \rightarrow \infty ),
\end{eqnarray}
where the term $(\alpha / \beta )^L$ should be neglected for consistency.
We observe that, for $\alpha$ and $\beta$ not to
close to one another, the travel times are of the order of $L$, as expected. 
Moreover, the average travel time $<T>$ is a function of both parameters $\alpha$ and $\beta$. As $\alpha \uparrow \beta$, the neglected correction 
terms $(\alpha/\beta)^L$ in  (\ref{d24}) and (\ref{d6}) become of ${\cal O}(1)$, and (\ref{d24}) is no longer valid.

For a large range of combinations of  $\alpha$ and $\beta$, the average travel times
have been measured with the simulation program, which is able to measure the travel
times of a variable set of individual (labeled) particles. The measured 
travel times are in excellent agreement with relation (\ref{d24}), even very close to the
transition line $(\alpha = \beta )$. In Figure 6a we show
a histogram of travel times in a system with
$L = 1000$, $\alpha = 0.50$, $\beta = 0.51$ in the low density phase
 near the transition line $\alpha = \beta $, where
travel times may vary considerably. The average
travel time for the specific measurement was 1049 time steps. Expression
(\ref{d24}) gives $<T> = 1049$, so the agreement is very good.

In the {\it high density phase} $(\alpha > \beta )$ the average travel time 
should be approximately $L/\beta$, because the speed in 
the bulk of the system is equal to $v_{J} = \beta$. The average travel time
$<T>$ can again be determined from definition (\ref{d23}) and (\ref{d8}) 
with the result
\begin{equation} \label{d25}
<T>  =  \frac{L}{\beta} - \frac{1-\alpha}{\alpha -\beta}
 \qquad (L \rightarrow \infty ).
\end{equation}
For $\alpha = 1$ the average travel time assumes the value $L/\beta$. For
$\alpha < 1$, the average travel time is smaller than $L/\beta$, because then
the particles travel with a velocity larger than $\beta$ just after 
entering the lattice. Comparison with the results of the low-density phase also 
shows that the statistical spread in travel times
is larger in the high-density phase, because the particles are often
blocked by other particles. Figure 6b  shows a histogram of the measured travel times for a lattice
of $L= 1000$ sites, and injection and removal rates of $\alpha = 0.51$ and
$\beta = 0.50$ respectively with a predicted average $<T>=1951$.
Obviously there is a large spread around the average travel time
(\ref{d24}).
The distribution is not symmetric and therefore not a Gaussian. The asymmetry
of the distributions in Figures 6a,b are related through particle-hole 
symmetry.

The average travel time $<T>$ on the transition line $\alpha = \beta$ can be
easily determined, using the average number of particles in (\ref{d23}).
The travel time combined with the flux
$j= \alpha / (1+\alpha )$ yields then
\begin{equation} \label{d26}
<T> = {<N>}/{j} = \textstyle{\frac{1}{2}} L \left( 1 + 
\frac{1}{\alpha} \right).
\end{equation}
This equation merely shows that in the coexistence region, the average
travel time is just the average of the low- and 
high-density travel times, because both phases are present and occupy
on average an  equal fraction of the system. 
The long time average $<T>$ agrees very well with (\ref{d26}).
Individual particles will have a considerably shorter
travel time of order $L$ when $R$ is located near the exit site, 
and a longer one of order $ L/\alpha$ when $R$ is 
located near the entrance site.

The prediction (\ref{d26}) is in excellent agreement with the
simulation results, as shown in Figure 7. We also note that the 
theoretical description of our totally asymmetric exclusion process 
in terms of occupation numbers
(indistinguishable particles) does not allow us to calculate 
travel times of labeled particles, or calculate the probability 
distributions in Figure 6. Such calculations would only be possible 
in a TASEP with labeled particles, as considered in 
Refs. \cite{evans,schreck}.

\section{TASEP on a ring with blockage site}

The fully synchronous TASEP of section 2.1 as a closed system with 
$N=\rho L$ particles 
and obeying periodic boundary conditions is a fully deterministic, 
rather uninteresting system in its NESS. 
For $\rho < \frac{1}{2}$ all sites are in pure free-flow
configurations and particles travel, say, counterclockwise with unit 
speed and flux $j=\rho$.
For $\rho > \frac{1}{2}$ all sites are in jammed configurations, and 
holes are traveling clockwise with unit speed, and the flux $j = 1-\rho$.

The dynamics becomes more interesting by inserting a stochastic 
blockage at site $i=L$ with a  transmission rate $\beta < 1$. The 
microdynamic equation for sites $i=2,3,\ldots ,L-1$ is the same as in 
(\ref{a1})-(\ref{a3}), but the fluxes referring to the blockage 
sites are
\begin{equation}\label{e1}
\hat{\jmath}_0 = \widehat{\jmath}_L = \widehat{\beta}\tau_L\sigma_1 ,
\end{equation}
where the Boolean variable $\hat{\beta}$ with expectation $< 
\hat{\beta}> =
\beta$, is defined in a similar manner as $\hat{\alpha}$ and 
$\hat{\beta}$ below (\ref{a3}). For $\beta = 1$ one recovers the fully
deterministic case with periodic boundary conditions.

First we observe that the dynamics at {\em fixed} $\beta$ is
invariant under the duality transformation
\begin{eqnarray} \label{e4}
\tau_i \leftrightarrow \sigma_{L-i+1} \nonumber \\
\rho \leftrightarrow 1 - \rho ,
\end{eqnarray}
and that the average occupation satisfies
\begin{equation} \label{e5}
<\tau_i> (\rho, \beta ) = < \sigma_{L-i+1}> (1-\rho , \beta ).
\end{equation}
A mean field theory for the bulk properties of this model in the
thermodynamic limit has already been given by Yukawa et al \cite{yukawa}, 
as well as extensive numerical simulations, specially for the 
coexisting phase region.
However, the correlation functions and density profiles have not yet been 
studied
analytically. In the Appendix  we have discussed the TASEP with open 
boundaries
and sublattice--parallel updating. The corresponding models on a ring 
with a single blockage site and sublattice--parallel updating
or  with random sequential updating  have also been solved exactly 
in Refs. \cite{joel-asep}--\cite{defect}.  
However, the analytic results show little similarity with 
those for the present model, and will not be discussed further.

The build up of dynamic correlations and structures may be analyzed 
in a
similar way as in section 3 for open systems, and one recovers 
(\ref{a12}),
{\em except} for site $i=L$. This implies that a cluster of holes 
upstream
of a particle cluster can only be created at the blockage site. Of 
course,
hole clusters upstream of particle clusters may be present in the 
initial
state anywhere on the lattice. In the low density phase, such 
configurations
will be destroyed in a time that is roughly equal to twice the size of the 
largest
hole cluster. Moreover, from a detailed analysis of the dynamics, 
similar
to section 3.1, one can derive exact relations for the 
microscopic 
dynamic correlations for the bulk of the low density phase, such as
\begin{eqnarray}\label{e6}
\tau_i \tau_{i+1} &=& 0 \qquad  {( \rm bulk} {\rm sites} \quad i \ll 
L - \lambda_B )  \nonumber \\
\tau_L \tau_1 &=& 0 \qquad        {( \mbox{\rm blockage site})} ,
\end{eqnarray}
where $\lambda_B$ is the width of the pile up region, downstream of
the blockage.

The continuity equation (\ref{a1}) and (\ref{e1}) yields then 
in  combination with (\ref{e6}) for the low density phase in the NESS,
\begin{eqnarray}\label{e7}
j = \beta <\tau_L > & = & <\tau_1> = <\tau_2> = \ldots = <\tau_i> \qquad 
(i \ll L-\lambda_B ) \nonumber \\
& = & <\tau_i> - < \tau_i \tau_{i+1}> \qquad (L-\lambda_B 
\stackrel{<}{\sim} i < L).
\end{eqnarray}
In the low density phase there is an excess density in the pile up 
region.
 Consequently, as the total density $\rho = N/L$ is fixed,
the density at bulk sites has the form
$<\tau_1> \simeq \rho \{1- {\cal O}(\lambda_B/L) \}$, as we
shall see later.

In fact, one can infer most of the results for the ring model with a
blockage from section 4, by considering the flux $j=\beta<\tau_L>$ 
across
 the link $(L,1)$ as the influx $j_0$ appearing in (\ref{a5}) for the open 
system.
This relation defines the effective input rate $\alpha_e$ through
the relation $j= \alpha_e (1-<\tau_1>)$ and yields in combination with 
$j=<\tau_1>$ in (\ref{e7}),
\begin{equation}
\alpha_e = j/(1-j)
\end{equation}
where $\alpha_e$ approaches $\rho /(1-\rho )$ in the thermodynamic 
limit.
The flux across the link $(L,1)$ can be equally considered as the 
outflux
$<\hat{\jmath}_L> = \beta_e <\tau_L>$ in (\ref{a5}) of the corresponding open system. This identifies the effective removal rate $\beta_e= \beta$ as the transmission
coefficient of the blockage site $L$.

The phase diagram for the system with a blockage can then be read off 
from
Figure 1, showing the flux $j(\rho )$ of the open system at a fixed
removal rate $\beta$. Consequently, for $\alpha_e < \beta_e$, or 
equivalently
for $j < \beta / (1+ \beta )$ (where $j \sim \rho$ for large 
systems), the
system is in the low density or free flow phase. If the density 
$\rho$ approaches 
$\rho_F = \beta / (1+\beta )$, or if $\alpha_e$ approaches $\beta$, 
then the
system enters the region of coexisting phases, and the pile up 
region, which had before a microscopic width
 $\lambda_B$ of approximate size 
$ 1/\ln (\beta_e / \alpha_e) = 1/\ln [\beta (1-\rho )/\rho ]$,
grows to macroscopic size, as in a wetting transition.
For $\rho > \rho_F$ an interface appears downstream of the blockage, 
at a location $R$, and the pile up region has the macroscopic 
size $L- R$. As the density $\rho$ increases further to $\rho_J = 
1/(1+\beta )$, the location $R$ moves further downstream, according to 
(see (\ref{a16}))
\begin{equation}\label{e9}
\rho = \left( \frac{R}{L}\right) \frac{\beta}{1+\beta} + 
\left( 1 - \frac{R}{L} \right) \frac{1}{1+\beta} 
\qquad (\rho_F < \rho < \rho_J ).
\end{equation}
As $\rho \uparrow \rho_J$, the free flow phase disappears $(R
\rightarrow 0)$, and the system goes into a pure jammed phase, 
where there is again a microscopic boundary layer, just 
upstream of the blockage with a deficit density.\\
In summary, the TASEP on a ring with a blockage site has the 
following phases,
\begin{itemize}
\item free flow phase: \hspace{17mm} $\rho < \rho_F = \beta / (1+\beta)$
\item coexisting phases: \hspace{12mm}  $\rho_F < \rho < \rho_J$ 
\item jammed phase: \hspace{16mm} $\rho > \rho_J = 1/(1+\beta )$. 
\end{itemize}
The above results were first obtained  and verified against computer  simulations by Yukawa et al. \cite{yukawa}.

The density profile $<\tau_i>$ in the free flow phase can be inferred 
from
the corresponding profile (\ref{d6}) for $\alpha_e < \beta_e$, 
and yields,
\begin{eqnarray}\label{e10}
<\tau_i> & = & j\{ 1 + \frac{1-\beta}{\beta} \zeta^{L-i}\} 
\quad (i < L) \nonumber \\
< \tau_L> & = & j/\beta,
\end{eqnarray}
where terms of ${\cal O}(\zeta^L)$ have been neglected. The relation above is valid for
\begin{equation}\label{e11}
\zeta \equiv \frac{\alpha_e}{\beta_e} = \frac{j}{\beta (1-j)} < 1 
\qquad {\rm or} \qquad  j< \rho_F.
\end{equation}
In fact, the first line in (\ref{e10}) also covers the case $i=L$. 
The relation between flux $ j(\rho) =
 <\tau_1>$ and density $\rho$ in the free flow phase follows by 
summing (\ref{e10}),
\begin{equation}\label{e12}
\rho = \frac{1}{L} \sum^{L}_{i=1} <\tau_i> = j \left\{ 1 + 
\frac{1}{L}
\frac{(1-\beta)(1-j)}{[\beta -j (1+\beta )]} \right\},
\end{equation}
where ${\cal O}(\zeta^L)$-terms have been neglected. The flux 
$j(\rho)$ can be solved from this quadratic equation, where the 
root with the minus sign is the physical root.
 For large 
systems $j$ differs only slightly from $\rho$. However, the 
${\cal O}(1/L)$-correction becomes more and 
more important as $j\uparrow\rho_F = \beta/(1+\beta )$ where the 
denominator in (\ref{e12}) diverges.
By a perturbation expansion to ${\cal O}(1/L)$
we find,
\begin{equation} \label{e13}
j(\rho) = \left\{ \begin{array}{ll}
\rho [ 1 - \left(\frac{1-\rho}{\rho_F - \rho} \right) 
\epsilon ] & (\rho < \rho_F) \\
\rho_F [ 1 - \left( \frac{1}{\rho -\rho_F} \right) \epsilon ] 
& ( \rho_F < \rho < \textstyle{\frac{1}{2}} ) 
\end{array} 
\right.
\end{equation}
with
\begin{equation} \label{e14}
\epsilon = \left( \frac{1-\beta}{1+\beta} \right) \frac{1}{L} 
\equiv \frac{ \Delta \rho}{L},
\end{equation}
where $\Delta \rho = \rho_J-\rho_F$ is the difference in density 
between the two coexisting phases. The numerical solution of (\ref{e12})
is plotted in Figure 8 as the dashed line.

We have again performed computer simulations on large and small 
systems to
test the density dependence of the flux $j(\rho )$. After 
preparing the system in a random initial configuration,
we let the system relax for $4 \times 10^4$ time step, 
after which it is assumed to be in the NESS. We have calculated 
time averages over $3 \times 10^5$ time steps, and ensemble 
averages over 50 different 
initializations. The agreement in the interval $ \rho < \rho_F $
between theory and simulations is 
excellent, even for
small $L$ and $\beta$, and for $\rho$ close to $\rho_F$ where 
the difference
between $j$ and $\rho$ is largest, as shown in Table 1 and Figure 8.
For $\rho_F < \rho < \textstyle{\frac{1}{2}} $  the difference between 
theory and simulations becomes somewhat larger. The reason is that
(\ref{e10})-(\ref{e12}) are only valid for $\zeta < 1$ or $j < \rho_F$.
As soon as $\zeta \simeq 1$ or $j \simeq \rho_F$  equation (\ref{e12})
starts to loose its validity because in (\ref{e10}) and (\ref{e12})
terms of $ {\cal O} (\zeta^L) \simeq {\cal O} (1)$ have been neglected.
Simulation results for $j(\rho )$, similar to those in Figure 8, have been presented in Ref. \cite{yukawa} without a theoretical explanation.

Next we consider the $n$-point correlations, which are given through 
(\ref{d5}) and (\ref{a13}), i.e.
\begin{equation} \label{e15}
< \tau_k \tau_{k+1} \ldots \tau_{k+n} > = (j/\beta ) (1-\beta )^n 
\zeta^{L-k},
\end{equation}
with $\zeta$ given in (\ref{e11}). 
This relation also gives the profile $<\tau_k > = j +$  $~< \tau_k 
\tau_{k+1}>$
of the pile up region downstream of the blockage. In Figure 9
the $n$-point correlations with $n=1,2$ have been compared with 
computer simulations and again there is good agreement.

Moreover, through arguments similar to those in section 5.1, we 
conclude
that the probability $P(k)$ to find the first site of the pile 
up region at site $k$ is proportional to the excess density 
$< \tau_k \tau_{k+1} >$, so that
$P(k) = \zeta^{L-k}/(1-\zeta)$. The average width $\lambda_B$ of the
blockage region and the fluctuation $\delta\lambda_B$ around this 
average are then 
found from (\ref{d63}) and (\ref{d64}) with $\zeta$ given through 
(\ref{e11}), i.e.
\begin{eqnarray}\label{e16}
\lambda_B &=& <(L-k)> = j/[\beta - j(1+\beta ) ] \nonumber \\
(\delta\lambda_B )^2 & = & < (L-k)^2> - <L-k>^2 \nonumber \\
&= &\beta j (1-j) /[\beta -j(1+\beta )]^2,
\end{eqnarray}
where $ <\ldots > = \sum_{k} (\ldots ) P(k)$.
Simulation results for $\lambda_B$ and $\delta\lambda_B$ have been 
presented in Ref. \cite{yukawa}.

For the {\em high density} phase all corresponding results can be
obtained from particle-hole symmetry. For instance, the profile of 
the
depletion region just upstream of the blockage site $L$ is given by
(\ref{d6}) with $\alpha_e = \beta$ and $\beta_e = j/(1-j)$ where $j(\rho , 
\beta )$ in the
jammed phase equals $j(1-\rho , \beta )$ in (\ref{e12}) 
in the free flow  phase.

The behavior of the interface in the {\em coexistence} region is very
different from that in the TASEP with {\em open} boundaries. In the
latter the location $R$ of the interface for a given $\alpha = 
\beta$, can be
anywhere on the lattice with equal probability, as the actual density 
fluctuates between $\beta / (1+\beta )$ and $1/(1+ \beta )$. In the
TASEP on the ring with blockage the density $\rho = N/L$ is fixed,
and $R$ is on average given through (\ref{e9}), and there are only small
fluctuations $\delta R$ around the average $R$. In summary, the 
previous
discussion confirms our intuitive interpretation that the TASEP on a 
ring with a
blockage behaves  the same as the TASEP
with open boundaries with injection rate $\alpha_e = j/(1-j)$ and 
removal rate $\beta_e = \beta$. This similarity holds for bulk properties, 
profiles and correlation functions.

\section{Conclusion}

In this paper, which is in part an account of Ref. \cite{tilstra}, we have
studied the non-equilibrium stationary state (NESS) of the totally asymmetric
exclusion process (TASEP) with fully synchronous and {\em deterministic}
$(p=1)$ bulk dynamics (i) for open systems, coupled to particle reservoirs
with injection rate $ \alpha$ and removal rate $\beta$, and (ii) for closed 
systems on a ring containing a stochastic blockage site with transmission 
rate $\beta$. Mean field theories and the Boltzmann equation give a totally 
inadequate description of these far-from-equilibrium states, because of the 
existence of strong short range correlations.

The theory presented here is based on two new ideas: (i) we derive,  
starting from the microdynamic equations for the TASEP, the explicit 
microscopic specifications of the configurations and order parameters
 for the separate phases; and (ii) we introduce an improved {\em 
mean field approximation} in (\ref{d2}) that neglects fourth and 
higher order correlation functions at the interface between the bulk phase 
and the boundary layers. The results for the profiles have been 
compared with extensive computer simulations, and turn out to 
be indistinguishable from the analytic results. We therefore 
{\em conjecture} that  our results for the open TASEP with 
 $\alpha \neq \beta$ are {\em exact} up to terms that are 
exponentially small in the system size $L$, for instance of order
$(\alpha/\beta)^L$ for $\alpha < \beta$.
Our results for the 
TASEP  on the ring with a blockage show small differences between 
theory and simulations. Clearly the identification of the flux through 
a blockage as both the influx (to define $\alpha_{e}$) {\it and} 
the outflux (to define $\beta_e$) of the open system is only approximate. 
Moreover, the neglected correction terms $(\alpha/\beta)^L$ start to
 become of ${\cal O}(1)$ as $\alpha_e \uparrow \beta_e$ or 
$ j \uparrow \rho_F$ (see Figure 8).

The first idea has enabled us to obtain exact results not only for bulk 
densities and currents, but also for the spatial and temporal correlation
functions. The second idea  has enabled us to obtain analytic results
 for the profiles in the boundary layers of density $<\tau_k>$ and cluster 
correlation functions $<\tau_k \tau_{k+1} \cdots>$. For the more general stochastic model with $p<1$ of \cite{rsss} no analytic results for profiles  and correlation functions are known.
 The ideas in 
(i) are akin to the elimination of the `Garden of Eden' states 
in \cite{rsss}, and those 
in (ii) to the `paradisical mean field approximation', hinted at in 
Ref. \cite{rsss}, but that lingo is not ours.

It is of interest to compare our results for fluxes and bulk densities
with known results. The phase diagram has been obtained before in
\cite{schuetz,hinrichsen} for the same TASEP with sublattice--parallel
updating, and in \cite{yukawa} for the fully synchronous TASEP on a ring 
with a blockage. It has a free flow phase $(\alpha < \beta )$,
and a congested phase $(\alpha < \beta)$, which coexist when $\alpha = \beta$.

In Ref. \cite{rsss} a more general {\em stochastic} TASEP with fully
synchronous dynamics has been analyzed, where particles hop  only
with probability  $p$ (with $p < 1)$. The corresponding phase diagram
contains the present $(p=1)$ phase diagram for $\alpha$ and $\beta$ less than
$\alpha_c \equiv 1-\sqrt{1-p})$, but it also contains more phases, such as the 
maximal current phase. In our $(p=1)$-model the maximal current
phase occurs only as a {\em bulk} phase for the special parameter values
$\alpha = \beta = 1$. In addition, the interface of finite width,
separating the coexisting phases for $\alpha = \beta <  1$, constitutes a
`microphase' of maximal current configurations. In the TASEP with $p=1$ 
the interface region (see section 4.2) contains only local configurations 
common to both the low {\it and} high density phase. These {\it common} 
local configurations are identical to the local configurations that 
constitute the maximal current phase. It would be interesting to find
 out for the fully synchronous TASEP with $p<1$ if it is possible to 
identify an interface region for $\alpha = \beta < \alpha_c$, which constitutes 
a `microphase of maximal current configurations' as well.

In the low density regime $(\alpha < \beta < 1-\sqrt{1-p})$ of Ref.
\cite{rsss} the flux and bulk density are found as
\begin{equation}\label{m1}
j= \alpha \frac{p-\alpha}{p-\alpha^2} \qquad \mbox{and} \qquad 
\rho = \frac{\alpha (1-\alpha )}{p-\alpha^2},
\end{equation}
which reduce for $p=1$ to the results in (16). The corresponding properties 
of the phase diagram in the high density regime can be obtained 
from particle-hole symmetry.

In order to illustrate the particle-hole attraction in these models, and its
dependence on the hopping rate $p$, we compare our results for the
deterministic version $(p=1)$ for the nearest neighbor correlation functions in the
bulk (for $i \ll L - \lambda_R$), i.e.
\begin{eqnarray}\label{m2}
<\tau_i \tau_{i+1}> & = & 0 \qquad ; \qquad < \sigma_i \sigma_{i+1}> = 1- 2\rho \quad
(i \; \epsilon \; \mbox{bulk}) \nonumber \\
<\tau_i \sigma_{i+1}> & = & <\sigma_i \tau_{i+1} > = j = \rho \quad (\forall i)
\end{eqnarray}
with those for the stochastic version $(p < 1)$ in Ref. \cite{rsss}, reading
\begin{eqnarray}\label{m3}
<\tau_i \tau_{i+1}> & = & \rho - j/p \qquad (i \; \epsilon \; 
\mbox{bulk}) \nonumber \\
<\sigma_i \sigma_{i+1}> & = & 1 - \rho - j/p 
\qquad (i \; \epsilon \; \mbox{bulk}) \nonumber \\
<\tau_i \sigma_{i+1}> & = & <\sigma_i \tau_{i+1}> = j/p \qquad (\forall i) ,
\end{eqnarray}
where relation (6) has been used. Of course (\ref{m3}) includes (\ref{m2})
for $p=1$.

We notice that in the low density phase of the deterministic version there
is a `hard core repulsion' for nearest neighbor sites, as $<\tau_i \tau_{i+1}>
= 0$ or equivalently, a strong attraction of particle-hole pairs on
nearest neighbor sites. In the stochastic version $(p <1)$ there is also a particle-hole 
attraction on nearest neighbor sites, because the covariance
\begin{equation}\label{m4}
<\tau_i\sigma_{i+1}> - <\tau_i> <\sigma_{i+1}> =
\frac{\alpha^2 (p-\alpha )^2}{p(p-\alpha^2 )} > 0,
\end{equation}
i.e. there exists a positive correlation between an occupied site and the
empty site, just in front of it, which increases monotonically 
as $p \uparrow 1$.

It would be very valuable to extend the present method, based on the 
ideas summarized in (i) and (ii) at the start of this section, to 
calculate the profiles, and spatial and temporal correlation functions
 in the stochastic TASEP of Ref. \cite{rsss} with its much richer 
phase diagram containing a maximal current phase.

\section*{Appendix A: Sublattice--Parallel Dynamics}
\setcounter{equation}{0}
\renewcommand{\theequation}{A.\arabic{equation}} 

In this appendix we will illustrate how the method of the present paper,
when  applied to the TASEP with sublattice -- parallel updating, yields the
exact results for the bulk properties and the asymptotic (large system)
results for profiles and correlation functions, as obtained in Refs. 
\cite{schuetz,hinrichsen}
The dynamics consists of two substeps. In the first step from $t 
\rightarrow t^\prime = t  + \frac{1}{2}$, the pairs $(1,L)$, (23), 
$\ldots 
(L-2,L-1)$ are updated in parallel where $L$ is even. There is only 
a possibly non-vanishing microscopic flux $\hat{\jmath}_i$ through 
the {\it even} link $(i,i+1)$, whereas the flux $\hat{\jmath}_i$
 through the {\it odd} link
vanishes.
 In the second step, from
$t \rightarrow t^{\prime\prime} = t^\prime  + \frac{1}{2} 
= t+1$, the pairs (1,2), (3,4), $\ldots 
(L-1,L)$ are updated in parallel, then $\hat{\jmath}_{i+1}$ is possibly 
non-vanishing and $\hat{\jmath}_i$ vanishes.

 If we denote $\tau_a (t), \tau_a(t+\frac{1}{2})$ and $\tau_a (t+1)$ 
respectively by $\tau_a, \tau^\prime_a$ and
$\tau^{\prime\prime}_{a}$ with $a = \{ 1,2,\ldots ,L \}$, then the
{\em microdynamic} equation for the first step $t \rightarrow t^\prime$
becomes:
\begin{equation}
\begin{array}{ll}
\tau^\prime_i  =  \tau_i \tau_{i+1} \qquad & ( 0 < i \leq L ; i \; 
\mbox{even}) \nonumber \\
\tau^\prime_{i+1} = \tau_{i+1} + \tau_i \sigma_{i+1} \qquad & (0 \leq i < L ;
i \; \mbox{even}),
\end{array}
\end{equation}
and for the second step $t^\prime \rightarrow  t+1$ 
\begin{equation}
\begin{array}{ll}
\tau^{\prime\prime}_{i} =  \tau^\prime_i + \tau^\prime_{i-1} \sigma^{\prime}_{i}
\qquad & (0 < i \leq L ; i \; \mbox{even}) \nonumber \\
\tau^{\prime\prime}_{i+1} = \tau^{\prime}_{i+1} \tau^{\prime}_{i+2} \qquad &
(0 \leq i < L ; i \; \mbox{even}) .
\end{array}
\end{equation}
With the conventions $\tau_0 = \widehat{\alpha}$ and $\sigma_{L+1} = \widehat{\beta}$,
as defined below (\ref{a3}), these equations also include the boundary
conditions for the open system. Moreover, we observe that the evolution
equations (A1)-(A2) are invariant under particle-hole exchange.

In the NESS there is again a constant site-independent flux through the
system. The flux $< \hat{\jmath}_i >$ out of {\em even} sites at integer times is
equal to the flux $<\hat{\jmath}^{\,\prime}_{i+1} >$ out of {\em odd} sites
at half-integer times, i.e.
\begin{eqnarray}
j &=& <\hat{\jmath}_i > = < \tau_i \sigma_{i+1} > \nonumber\\
& = & < \hat{\jmath}^{\,\prime}_{j+1}> = < (\tau_{i+1} + \tau_i \sigma_{i+1})
(\sigma_{i+2} + \tau_{i+2} \sigma_{i+3}) > ,
\end{eqnarray}
where $\tau_0 = \widehat{\alpha}$ and $\sigma_L = \widehat{\beta}$ and $i$ is 
{\em even}.

By studying the dynamics of clusters as in section 3 one finds that (A1)-(A2)
impose some very strong constraints on the allowed configurations, i.e.
\begin{eqnarray}
&(\tau\sigma )^{\prime\prime}_{i-1,i}  = \tau^{\prime}_{i-1} 
\tau^{\prime}_{i} \sigma^{\prime}_{i-1} \sigma^{\prime}_{i} = 0 &\nonumber \\
&(\tau\tau\sigma\sigma )^{\prime\prime}_{i-1,i+2}  =  \tau^{\prime}_{i-1}
\tau^{\prime}_{i} \sigma^{\prime}_{i+1} \sigma^{\prime}_{i+2} = & \nonumber\\
& \{ (\tau\tau\tau )_{i-1,i+1} + (\tau\sigma\tau\tau )_{i-2,i+1} \}
\sigma_i \sigma_{i+1} \tau_{i+2} \tau_{i+3} = 0 & .
\end{eqnarray}
This implies that configurations containing $(\ldots 10_+ \ldots )$ and
$(\ldots 1100_+ \ldots )$ can never be created at integer times
if one starts from an empty initial state. Such configurations 
are therefore {\em absent} in the NESS for {\em any} value of $\alpha$ and
$\beta$. The subscripts $(\pm )$ on $\tau_{\pm}$ indicate that the
relevant site has  an {\em even} (+) or an {\em odd} $(-)$ label.
Of course configurations $(\ldots 10_- \ldots )$ as well as $(\ldots 0011 
\ldots )$ are allowed at integer times.

Moreover, by arguments similar to those in section 3.1 and 3.2, one shows again
that the first particle-cluster can only be created at the exit site.
Let $k_0$ be the location of the last particle on the last cluster
of particles, then the interval downstream of $k_0$ contains only 
{\em isolated particles}, separated by holes (free flow configurations with
interval density $\rho (< k_0) < \frac{1}{2})$, and that upstream
of $k_0$ contains only {\em isolated holes} (jammed configurations with 
interval density $\rho (> k_0 ) > \frac{1}{2}$). All conclusions in the last
three paragraphs of section 3.1 carry over to the TASEP with sublattice --
parallel dynamics, and so does the phase diagram.

Next, we consider the bulk properties for large systems $(L \rightarrow
\infty )$ in the free flow phase $(\alpha < \beta $ and $\rho_F <
\frac{1}{2}$). The low density phase is again characterized by the 
microscopic order parameter $\tau_i \tau_{i+1} = 0$. To perform these calculations,
we start, as in section 4.1, from the constant flux relations (A3) in combination with
the vanishing order parameter,
\begin{eqnarray}
j & = & <\hat{\jmath}_i >  = <\tau_2 > = \ldots  = < \tau_i> = 
\beta < \tau_L > \nonumber \\
& = & < \hat{\jmath}\,^{\prime}_{i+1}>  = \alpha (1-<\tau_1 > )  
=  <\tau_2 + \tau_3 > =  \nonumber \\
&=& \ldots = < \tau_i + \tau_{i+1} >
\end{eqnarray}
valid for $i \ll L - \lambda_R$ and $i$ even. This implies for 
{\em odd bulk} sites $<\tau_- > = 0$, and for {\em even bulk} 
sites $<\tau_+ > = \alpha$ and $<\tau_L > = \alpha /\beta$. 
If $<j_+>$ denotes the flux out of even bulk sites, and $<j_->$ the 
one out of odd sites, then 
the results for the free flow phase $(\alpha < \beta)$ can be summarized as,
\begin{eqnarray} 
<\tau_+> =  \alpha & \qquad <\tau_-> = 0  &\qquad <\tau_L>= 
\alpha/\beta  \nonumber \\
<\hat{\jmath}_+> = \alpha & \qquad <j_-> = 0 & .
\end{eqnarray}
The corresponding relations for the
jammed phase $(\alpha > \beta )$ can be obtained from the relations 
(\ref{a7})  for  particle-hole symmetry, and read:
\begin{eqnarray}
 <\tau_+> (\alpha,\beta) &= 1 - <\tau_-> (\beta,\alpha) &=1 \nonumber \\
<\tau_-> (\alpha,\beta) &= 1 - <\tau_+> (\beta,\alpha) &=1- \beta \nonumber \\
<\tau_1> (\alpha,\beta) &= 1 - <\tau_L> (\beta,\alpha) &=1-\beta/ \alpha, 
\end{eqnarray}
and for the fluxes, using  (\ref{a8}),
\begin{eqnarray}
<\hat{\jmath}_+> (\alpha,\beta) &= &<\hat{\jmath}_+> (\beta,\alpha)
 = \beta \nonumber \\
<\hat{\jmath}_{-}> (\alpha,\beta)  &=& 0 .
\end{eqnarray}
Next, we consider the profile in the right boundary layer of the {\em low
density} phase, and we construct the dynamics of the cluster functions $T$, 
as in (\ref{a10}). By specializing these equations to the low density phase
one arrives after lengthy, but straightforward algebra, at a coupled
hierarchy of equations for the correlation functions.

Let $i$ or $i+1$ (with $i$ = even) be the last particle position on
the last particle cluster, then we find in the NESS the exact relations,

\begin{eqnarray}
<T_{ik}> & = & <T_{i,k-1}> = < T_{i,k+1} > + \nonumber \\
&& <(\tau\sigma\sigma T)_{i-2,k+1}> + <(\tau\sigma \tau\sigma T)_{i-2,k+1}> 
\nonumber \\
< T_{i+1,k} > & = & < T_{i+1,k-1}> = <T_{i+1,k+1}> + <(\tau\sigma T)_{i,k+1} > ,
\end{eqnarray}
where $i$ and $k$ are both {\em even}. The present set of coupled equations is
the analog of (\ref{d1}). By application of the mean field assumption, 
formulated in (\ref{d1})-(\ref{d2}), the above set of equations simplifies to
the  set of recursion relations with $i$ and $k$ even,
\begin{eqnarray}
<T_{ik}> & = & <T_{i,k+1}> + \alpha < T_{i+1,k+1}> + \alpha^2 <T_{i+2,k+1} >
\nonumber \\
< T_{i+1,k}> & = & <T_{i+1,k+1} > + \alpha <T_{i+2,k+1}> .
\end{eqnarray}
The boundary condition for this set is included by setting $k=L$. 

Our special
mean field assumption for the {\em low density} phase neglects again higher order
correlations between on the one hand the particle-hole pairs $\tau\sigma$ and 
$\tau\sigma\tau\sigma$, and on the other hand the tailing particle cluster 
at the interface of bulk phase and boundary layer. The solution of
these recursion relations yields 
\begin{eqnarray}
<T_{ik}> & = & <T_{i,k-1}> = \frac{\alpha}{\beta} <T_{i+1,k}> =
\frac{\alpha}{\beta} < T_{i+1,k+1}> \nonumber \\
& = & \left( \frac{\alpha}{\beta} \right)^{L-i+1} 
(1-\beta )^{\frac{1}{2}(k-i)} .
\end{eqnarray}
From these results and from (\ref{a4a}) we obtain the density profile of the
boundary layer near the exit for {\em even} sites $i$,
\begin{equation}
<\tau_i> = \alpha + (1-\beta ) \left( \frac{\alpha}{\beta}\right)^{L-i+1} .
\end{equation}
For {\em odd} sites $i+1$ follows similarly,
\begin{equation}
<\tau_{i+1}> = (1-\beta ) \left( \frac{\alpha}{\beta}\right)^{L-i} .
\end{equation}
The results for the profiles and correlation functions (A11)-(A13) are in full
agreement with the {\em exact} results of Refs. \cite{schuetz,hinrichsen}
 for large systems,
when terms of order $(\alpha / \beta )^L$ have been neglected.

In close parallel to section 5.2, we may also calculate the spatial and
temporal correlation functions in the bulk of the low density phase.
By setting the microscopic order parameter $\tau_i \tau_{i+1} = 0$
in (A1)-(A2), the microdynamic equation for bulk sites $(i \ll L - \lambda_R)$
reduces to
\begin{equation}
\tau_i(t+1) = \tau_{i-2}(t) \quad ; \quad \tau_{i+1} (t) = 0 ,
\end{equation}
where $i$ is even. The correlation function in the NESS with
$i$ and $R$ even are then,
\begin{eqnarray}
< \tau_{i+R}(t) \tau_i (0) > & = & <\tau_{i+R-2t} (0) \tau_i (0) > \nonumber \\
& = & < \tau_{i+R}(t) > <\tau_i > = \alpha^2 ,
\end{eqnarray}
holding for $0 \leq \{ i,j\} << L - \lambda_R$ with $j=i+R-t$ and 
$R \neq 2t$. For $i$ and/or
$R$ odd, the correlation function vanish, as $<\tau> = 0$ for {\em odd}
sites. We observe that the occupations between two even sites in the 
low density
phase are {\em uncorrelated}, as a consequence of the {\em sublattice -- parallel} dynamics.  
In the corresponding case of fully parallel dynamics, the
occupations are {\em correlated}, as is shown in (\ref{d16}) and (\ref{d18}).
The absence of correlation is understandable here, as a particle,
attempting
to enter the system in the low density phase at site 1 is never blocked.

\section*{Acknowledgements}
One of us (M.H.E.) wants to thank H. Knops for stimulating discussions 
and A. Schadschneider for clarifying correspondence.

\newpage

\section*{Figure Captions}

\noindent
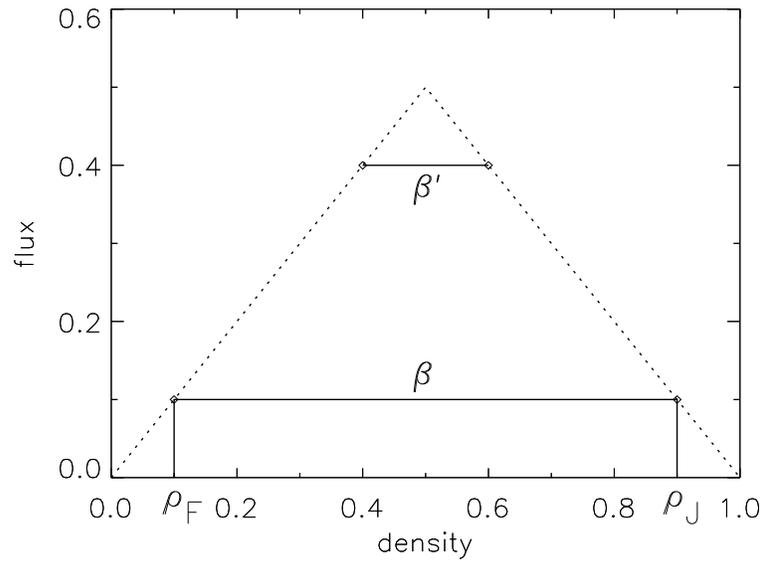
Figure 1: Flux $j$ as a function of density $\rho$ at constant 
removal rate 
$\beta$ with $\beta < \beta^\prime$. For $\rho_F < \rho < \rho_J$ 
with 
$\rho_F = \beta /(1+\beta )$ and $\rho_J = 1/(1+\beta )$ the flux
$j = \rho_F$ remains constant. The triangle bounding the region 
$j < \min \{ \rho , 1-\rho \}$ 
of coexisting phases is called the fundamental diagram in traffic 
flow problems.
\\[2ex]
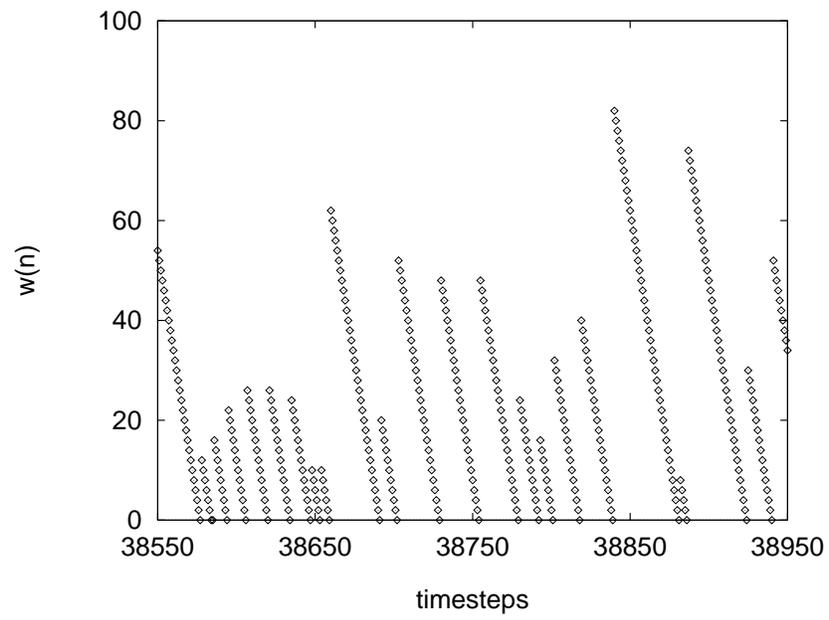
Figure 2: Instantaneous interface width $w$ between coexisting phases 
as a function of time, measured for $L = 1000$, $\alpha = \beta = 0.9$. 
Notice the saw-tooth behavior of the width,
corresponding to right and/or left jumps of the interface boundaries.
\\[2ex]
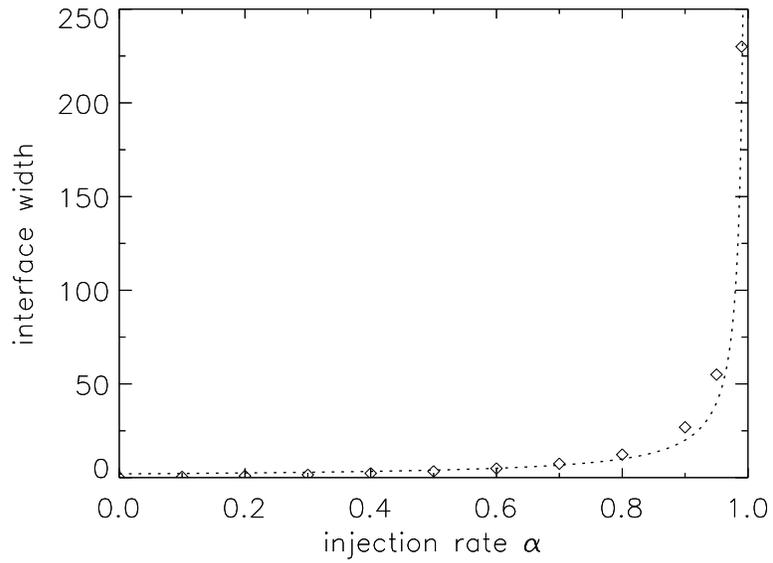
Figure 3: Average interface width $< w >$ between coexisting phases, 
measured over an interval of $2 \times 10^4$ time steps, as a function 
of injection rate $\alpha$, in a system with $L=1000$, and compared with 
the theoretical prediction $< w > = 2 /(1-\alpha )$. As $\alpha 
\uparrow 1$, the width of the interface changes from microscopic 
to macroscopic.
\\[2ex]
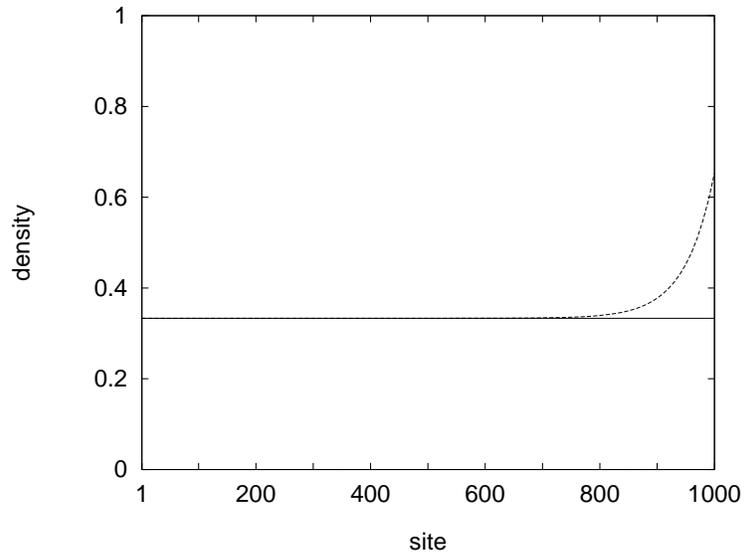
Figure 4: (a) Density profiles for $\{ \alpha = 0.50; \beta =1 \}$ and
$\{ \alpha = 0.50$ and $\beta = 0.51 \}$ for a system with $L=1000$. 
The bulk densities are the same (see (16)), except in the boundary layer, 
where the excess density is $<\tau_i \tau_{i+1}>$, averaged over $1.3 \times
10^7$ time steps.
(b) The 3-point correlation
$< \tau_i \tau_{i+1} \tau_{i+2}>$. In both plots the theoretical and
simulation results are indistinguishable. The latter are time 
averaged over $1.8\times 10^4$ time steps.
\\[2ex]
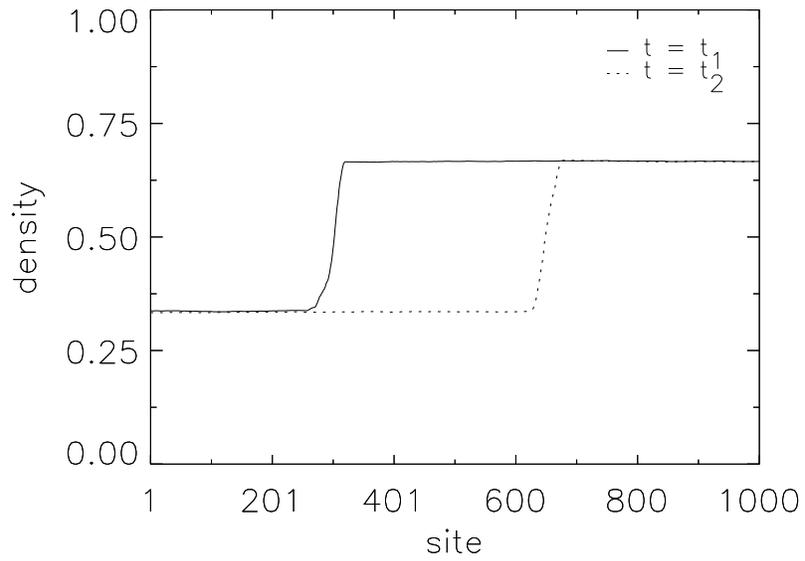
Figure 5: Density profiles in the coexistence region, time averaged 
over 4000 time steps. At time $t_1 /t_2$ the interface is at $R$ is 
302/649, and the average number of particles $<N>$ is 566/450 respectively.
\\[2ex]
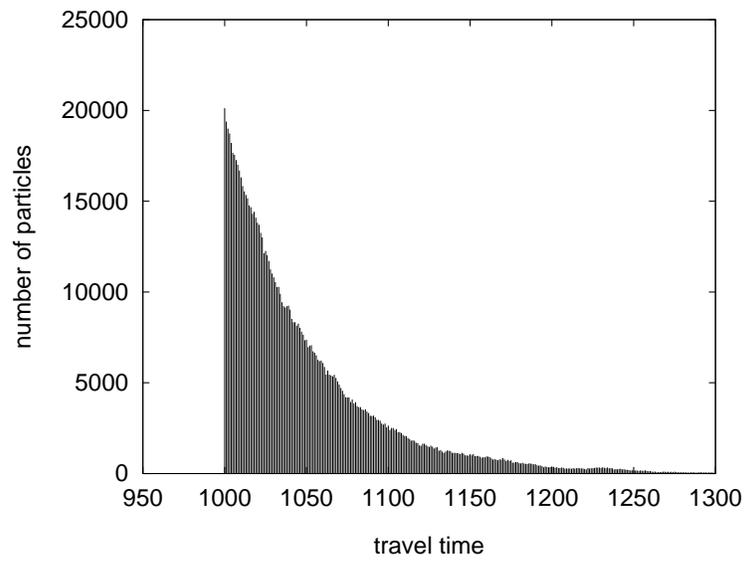
Figure 6: Histogram of travel times for a system of $L=1000$ (a) in 
the
low density phase with $\{ \alpha = 0.50; \beta = 0.51\}$ and (b) in 
the high
density phase with $\{ \alpha = 0.51 ; \beta = 0.50 \}$.
\\[2ex]
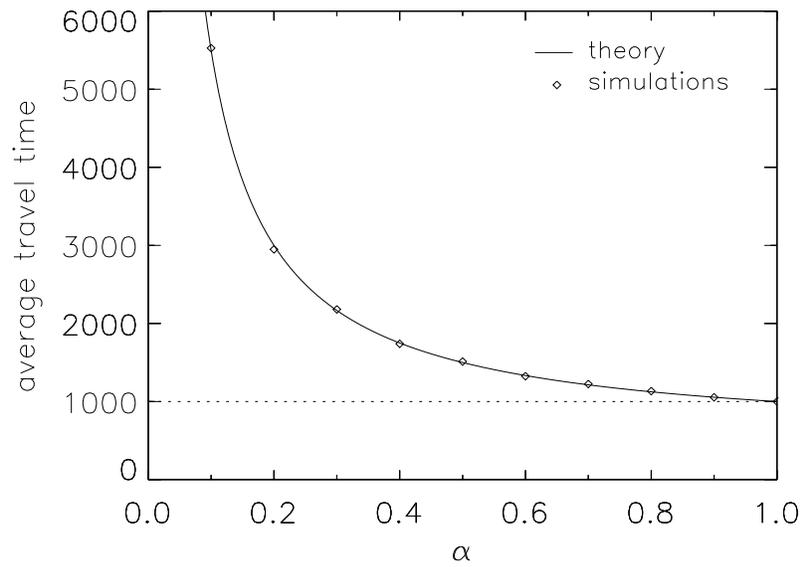
Figure 7: Average travel time $< T >$ as a function of $\alpha = \beta$ 
for coexisting phases in a system with $L=1000$ sites.
\\[2ex]
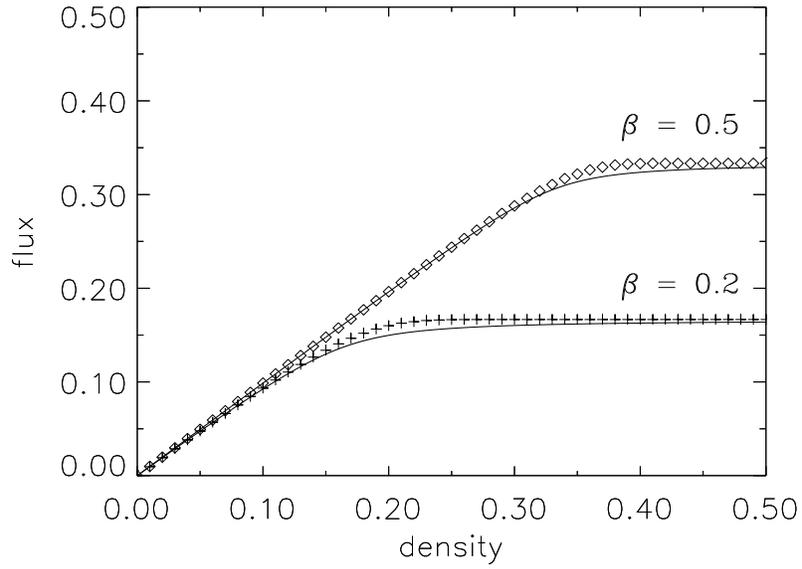
Figure 8: `Equation of state' for the flux $j(\rho)$ from
 (\ref{e12}) and (\ref{e13}) (dashed line) compared with simulation 
results for $L=1000$ sites at $\beta =0.2 (\rho_F \simeq 0.17)$ and
$\beta =0.5 (\rho_F \simeq 0.33)$.
The smooth crossover at $\rho =\rho_F$ 
is derived from the profile (\ref{e10}) of the blockage region.
\\[2ex]
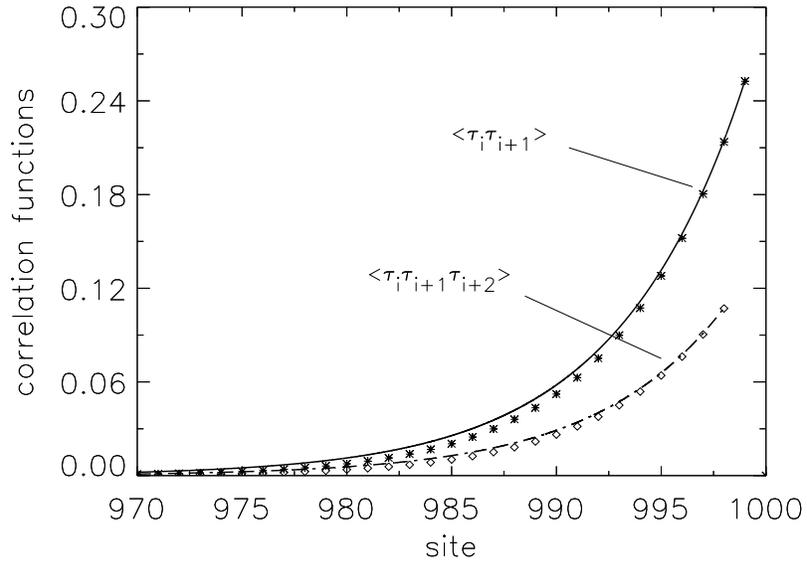
Figure 9: Correlation function $< \tau_i \tau_{i+1} \tau_{i+2}>$ as 
a function of $i$. The points are simulation results; the curve 
represents the relation (\ref{e15}), with parameters $L=1000, \beta = 0.5$
and $\rho = 0.3 $.
\\[2ex]

\section*{Table Captions}

\noindent
Table 1: Equation of state $j(\rho,\beta)$.

\begin{table}[hbt]
\[
\begin{array} {|c|c|c|c|}\hline                                                

        & L=100 & \beta = 0.25 & \rho_F = 1/5\\ \hline
N         & \rho & j(sim) & j(theor) \\ \hline 
10         & 0.10  & 0.0955        & 0.0958   \\
15 & 0,15  & 0.1401        & 0.1384   \\
18         & 0.18  & 0.1638        & 0.1599  \\ \hline
     & L = 100 & \beta = 0.5& \rho_F = 1/3 \\ \hline
20       & 0.2   & 0.1964        & 0.1962 \\
30   & 0.3   & 0.2882        & 0.2857 \\ \hline
      & L = 1000 & \beta = 0.5 & \rho_F = 1/3 \\ \hline
300    & 0.30  & 0.2981        & 0.2980 \\ 
330 & 0.33  & 0.3243        & 0.3230 \\ \hline
      & L = 100 & \beta = 0.75 & \rho_F = 3/7 \\ \hline               
30      & 0.30  & 0.2971        & 0.2977 \\
40   & 0.40  & 0.3914        & 0.3910 \\ \hline
      & L = 1000 & \beta = 0.1 & \rho_F = 1/11 \\ \hline
60    & 0.60 & 0.0587         & 0.0586 \\
85   & 0.85  & 0.0806        & 0.0797 \\ \hline
\end{array}
\]
\end{table}
\end{document}